\documentclass{aa} 

\usepackage{multirow}
\usepackage{graphicx}
\usepackage{hyperref} 
\hypersetup{colorlinks=true,urlcolor=blue, citecolor=blue}
\def\sun{\ensuremath\odot}
\graphicspath{{Figures/}}
\usepackage{txfonts}

\newcommand{\msun}{\ensuremath{M_{\odot}}}   
\newcommand{\nuc}[2]{\ensuremath{{}^{#2} \textrm{#1}}}

\begin{document} 

 \title{Grids of stellar models with rotation} 
 \subtitle{VIII. Models from 1.7 to 500\,\msun\ at metallicity $Z = 10^{-5}$}
 \titlerunning{Grids of stellar models with rotation at $Z = 10^{-5}$}
 \authorrunning{Sibony et al.}

 \author{Yves Sibony\inst{1}, Kendall G. Shepherd\inst{2}, Norhasliza Yusof\inst{3}, Raphael Hirschi\inst{4,5}, Caitlan Chambers\inst{4}, Sophie Tsiatsiou\inst{1}, Devesh Nandal\inst{1}, Luca Sciarini\inst{1}, Facundo D. Moyano \inst{1}, Jérôme Bétrisey\inst{1}, Gaël Buldgen\inst{1,6} Cyril Georgy\inst{1}, Sylvia Ekström\inst{1}, Patrick Eggenberger\inst{1}, Georges Meynet\inst{1}} 

 \institute{Observatoire de Genève,
              Chemin Pegasi 51, 1290 Versoix, Switzerland\\
              \email{yves.sibony@unige.ch}
              \and SISSA, via Bonomea 365, 34136 Trieste, Italy
              \and Department of Physics, Faculty of Science, Universiti Malaya, 50603 Kuala Lumpur, Malaysia
              \and Astrophysics Group, Keele University, Keele, Staffordshire ST5 5BG, UK
              \and Kavli Institute for the Physics and Mathematics of the Universe (WPI), University of Tokyo, 5-1-5 Kashiwanoha, Kashiwa 277-8583, Japan
              \and STAR Institute, Université de Liège, Liège, Belgium}

 \date{}

 \abstract
 {Grids of stellar evolution models with rotation using the Geneva stellar evolution code (\textsc{Genec}) have been published for a wide range of metallicities.}
 {We introduce the last remaining grid of \textsc{Genec} models, with a metallicity of $Z=10^{-5}$. We study the impact of this extremely metal-poor initial composition on various aspects of stellar evolution, and compare it to the results from previous grids at other metallicities. We provide electronic tables that can be used to interpolate between stellar evolution tracks and for population synthesis.}
 {Using the same physics as in the previous papers of this series, we computed a grid of stellar evolution models with \textsc{Genec} spanning masses between 1.7 and 500\,$M_\odot$, with and without rotation, at a metallicity of $Z=10^{-5}$.}
 {Due to the extremely low metallicity of the models, mass-loss processes are negligible for all except the most massive stars. For most properties (such as evolutionary tracks in the Hertzsprung-Russell diagram, lifetimes, and final fates), the present models fit neatly between those previously computed at surrounding metallicities. However, specific to this metallicity is the very large production of primary nitrogen in moderately rotating stars, which is linked to the interplay between the hydrogen- and helium-burning regions.}
 {The stars in the present grid are interesting candidates as sources of nitrogen-enrichment in the early Universe. Indeed, they may have formed very early on from material previously enriched by the massive short-lived Population III stars, and as such constitute a very important piece in the puzzle that is the history of the Universe.}


 \maketitle

\section{Introduction}

Libraries of stellar models computed for different metallicities with otherwise identical input physics are useful for many purposes, going from population synthesis to the chemical and photometric evolution of galaxies \citep[see e.g.][to name a few such libraries]{Bressan2012,Chieffi2013,Hidalgo2018}. Here, we present a grid of stellar models for a very low metallicity given by a mass fraction of heavy elements $Z$ equal to 10$^{-5}$ (sum of the mass fractions of the elements heavier than helium). These models, computed with the Geneva stellar evolution code (\textsc{Genec}), allow us to explore the effects of changing the initial mass and rotation at that specific metallicity. Comparisons with the other \textsc{Genec} grids of this series \citep{Z0142012, Z0022013, Z000042019, Pop32021, Z0062021, Yusof2022} allow exploration of the impact of changing the metallicity while keeping the remaining physical ingredients the same.

Although the low-metallicity models that we present in this work cannot be compared directly to observed individual stars, this new library of models is useful for studying the integrated properties of stellar populations, such as ionising fluxes, photometric magnitudes, and so on. Also, as will be discussed in forthcoming papers, such models predict stellar yields, which are useful for chemical evolution models.

We chose this metallicity of 10$^{-5}$ (which corresponds to ${\rm [Fe/H]}\sim-3$) because it marks the entry into the domain of the extremely metal-poor (EMP) stars  \citep{BC2005}. This is a metallicity domain that has been less explored than others, such as $Z=0$ (Population (Pop) III) or the low-$Z$ Magellanic Clouds. Among the published works presenting grids of models at that metallicity, we can cite \citet{Herwig2004}, \citet{GilPons2013}, \citet{Limongi2018}, \citet{GilPons2021}, and \citet{Ventura2021}. The present paper is, to our knowledge, the first to offer a grid of both non-rotating and rotating models spanning such an extended range of masses (from 1.7 to 500\,$M_\odot$).

We might wonder whether or not such models would present large differences with the zero-metallicity stellar models and what they would bring as new features with respect to Pop~III stellar models. As we show below \citep[see also][]{Tsiatsiou2024}, even a small amount of metal brings huge differences. One example is the consequence that metal can have on the primary nitrogen production \citep{POPIII2008}. Indeed, adding a small amount of metal decreases the temperature at which hydrogen is transformed into helium and this has an impact on the ease with which mixing can occur between the H- and He-burning regions. We come back to this point below.

This paper is organised as follows:
In Sect.~\ref{Ingredients}, we briefly describe the main physical ingredients of the stellar models. The properties of these models are discussed in Sect.~\ref{Properties}.
The comparison between the $Z=10^{-5}$ models and the models previously computed at other metallicities is the topic of Sect.~\ref{diffZ}. Some comparisons with stellar models from other works are given in Sect.~\ref{comparisons}. We present caveats and synthesise the main conclusions of our work in Sect.~\ref{Discussion_conclusion}.

\section{Stellar models: Physical ingredients and electronic tables}
\label{Ingredients}

We used the Geneva stellar evolution code (\textsc{Genec}) to compute a grid of stellar models at an initial metallicity of $Z=10^{-5}$. These models have the following initial masses: 1.7, 2, 2.5, 3, 4, 5, 7, 9, 12, 15, 20, 25, 30, 40, 60, 85, 120, 150, 200, 300, and 500\,$M_\odot$. For each mass, we compute both a non-rotating and a rotating model. We evolve the models as far as possible, typically up to the helium flash (stars with initial mass $M_{\rm ini}\leq2\,M_\odot$), the early asymptotic giant branch (AGB) phase ($2.5\,M_\odot\leq M_{\rm ini}\leq7\,M_\odot$), and at least the end of core carbon-burning ($M_{\rm ini}\geq9\,M_\odot$).

The present models are based on the same physics as what has been used in the series of models published by \citet{Z0142012}, \citet{Z0022013}, \citet{Z000042019}, \citet{Pop32021}, \citet{Z0062021}, and \citet{Yusof2022}. The interested reader can refer in particular to 
\citet{Z0142012}, \citet{Georgy2012}, and \citet{Z0022013}, where detailed descriptions of various physical ingredients of the models are presented. We restrain here to a short reminder of the physics of mass loss, convection, and rotation.

We use a metallicity-dependent mass-loss rate $\dot{M}$ during the main sequence (MS), blue and yellow supergiant (BSG and YSG, between which the cut-off temperature is $\log{(T_{\rm eff}\,[K])}=3.9$), and Wolf-Rayet (WR) phases, scaling as $\dot{M} = (Z/Z_\odot)^\alpha \dot{M}_\odot$, with $\alpha=0.85$ \citep[MS and BSG, from][]{Vink2001}, $\alpha=0.5$ \citep[YSG, from][]{DeJager1988}, or $\alpha=0.66$ \citep[WR, from][we note that this stage is not reached by any star in the present grid with such a low metallicity]{Eldridge2006}. In advanced phases with cooler ($\log{(T_{\rm eff}\,[K])}<3.7$) effective temperatures, we use a metallicity-independent mass-loss rate which depends only on the luminosity of the star.\\
For stars with a high Eddington factors, new mass-loss rates have been proposed by e.g. \citet{Bestenlehner2020} and \citet{SanderVink2020}. We nevertheless use the same recipe for stars in this situation as for the lower-mass ones because of our concern for homogeneity in the input physics across \textsc{Genec} grids at all metallicities. Finally, it is possible for a rotating star to reach the critical limit, where the centrifugal acceleration at the equator balances gravity such that the effective gravity is $0$. In that case, mass can be lost in a process we call mechanical mass loss. The details of the physics and of their implementation in \textsc{Genec} are complex, and we refer the reader to Sect.~2 of \citet{Georgy2013} for a thorough description of the process.

The present models have been computed using the Schwarzschild criterion for determining the size of convective zones. The 
convective core radius is obtained by $R_{\rm{CC}}=R_{\rm{Sch}}+0.1H_p$, where $R_{\rm Sch}$ is the radius obtained by applying the Schwarzschild criterion for convective instability, and $H_p$ is the pressure scale height at the Schwarzschild boundary. We apply such a step overshoot only during the core H- and He-burning phases. In the overshooting region, we assume, as in the rest of the core, that convection is adiabatic. No overshooting is applied on intermediate convective zones, nor undershooting on convective envelopes.

Rotating stars are initialised such that they reach a surface rotation velocity of 40\,\% the critical velocity ($\upsilon_{\rm ini}/\upsilon_{\rm crit}=0.4$) when three thousandths of hydrogen mass fraction have been transformed into helium at the centre. For the mixing by rotation we consider here models without a magnetic field (either interior or at the surface). We account for the transport of the angular momentum and of the chemical species by meridional currents and shear instabilities as described in \citet{Zahn1992} using the vertical and horizontal shear turbulence expressions given by respectively
\citet{Maeder1997} and \citet{Zahn1992}. As a result, rotation-induced mixing (apart from overshooting) is the only mixing mechanism present in radiative zones.

We use an alpha-enhanced initial chemical composition. The abundances in $\log{({\rm X/H})}+12$ and in mass fraction of the main isotopes are given in Table~\ref{tab:initial_composition}.

\begin{table}[h!]
    \caption{Initial abundances of the models for the main elements included in the reaction networks.}
    \centering
    \begin{tabular}{rr|rr}
    \hline\hline
    Element & Isotope & $\log{({\rm X/H})}+12$ & mass fraction\\
    \hline
    H & $^{1}$H & 12.00 & 7.516e-01 \\
    &&& \\
    He & & 10.92 & 2.484e-01 \\
    & $^{3}$He & & 4.123e-05 \\
    & $^{4}$He & & 2.484e-01 \\
    &&& \\
    C & & 5.16 & 1.296e-06 \\
    & $^{12}$C & & 1.292e-06 \\
    & $^{13}$C & & 4.297e-09 \\
    &&& \\
    N & & 3.99 & 1.026e-07 \\
    & $^{14}$N & & 1.022e-07 \\
    & $^{15}$N & & 4.024e-10 \\
    &&& \\
    O & & 5.75 & 6.823e-06 \\
    & $^{16}$O & & 6.821e-06 \\
    & $^{17}$O & & 3.514e-10 \\
    & $^{18}$O & & 2.001e-09 \\
    &&& \\
    Ne & & 4.80 & 9.448e-07 \\
    & $^{20}$Ne & & 9.205e-07 \\
    & $^{21}$Ne & & 7.327e-10 \\
    & $^{22}$Ne & & 2.354e-08 \\
    &&& \\
    Na & $^{23}$Na & 2.38 & 4.135e-09 \\
    &&& \\
    Mg & & 4.09 & 2.233e-07 \\
    & $^{24}$Mg & & 2.012e-07 \\
    & $^{25}$Mg & & 1.030e-08 \\
    & $^{26}$Mg & & 1.179e-08 \\
    &&& \\
    Al & $^{27}$Al & 2.58 & 7.694e-09 \\
    &&& \\
    Si & $^{28}$Si & 3.99 & 2.060e-07 \\
    \hline
    \end{tabular}
    \label{tab:initial_composition}
\end{table}

We use the same nuclear reaction rates as for all previous grids of this series. Most of them are taken from the NACRE database \citep{Angulo1999}, with the $\nuc{N}{14}(p,\gamma)\nuc{O}{15}$ rate from \citet{Mukhamedzhanov2003}, the $3\alpha$ rate from \citet{fynbo2005}, the $\nuc{C}{12}(\alpha,\gamma)\nuc{O}{16}$ rate from \citet{Kunz2002}, and the $\nuc{Ne}{22}(\alpha,n)\nuc{Mg}{25}$ rate from \citet{Jaeger2001}. For stars with initial mass $M_{\rm ini}\leq7\,M_\odot$, the NeNa and MgAl cycles are not computed explicitly. For stars with $M_{\rm ini}\geq9\,M_\odot$, these cycles are computed explicitly, with the $\nuc{Ne}{21}(p,\gamma)\nuc{Na}{22}$ rate from \citet{Iliadis2001}, the $\nuc{Ne}{22}(p,\gamma)\nuc{Na}{23}$ rate from \citet{Hale2001}, and the other reaction rates from \citet{Angulo1999}.

Electronic tables giving properties of the models along their evolution are publicly accessible\footnote{\url{https://obswww.unige.ch/Research/evol/tables_grids2011/}}. For each star, a table contains 400 lines where each line corresponds to a certain moment of the evolution. In other words, points that share the same number in tracks of different stars correspond to the same evolutionary stage. This facilitates interpolation between different tracks and population synthesis endeavours.

\section{Properties of the stellar models}
\label{Properties}

Table~\ref{TabListModels} shows a few properties of the present models including the very massive stars (VMSs), on the zero-age main sequence (ZAMS) and at the end of the core hydrogen, helium, and carbon (if applicable) burning phases. In Sect.~\ref{Evolution_surface} and~\ref{Evolution_centre} we focus on stars between 1.7 and 120\,$M_\odot$, and dedicate Sect.~\ref{VMS} to the study of the VMSs.

\subsection{Evolution of surface properties}
\label{Evolution_surface}
\begin{table*}[h!]
\caption{Properties of the stellar models on the ZAMS and at the end of the core hydrogen, helium, and carbon burning phases.}
\centering
\scalebox{.65}{\begin{tabular}{|cccc|cccccc|cccccc|cccccc|}
\hline\hline
\multicolumn{4}{|c|}{} & \multicolumn{6}{c|}{End of H-burning} & \multicolumn{6}{c|}{End of He-burning} & \multicolumn{6}{c|}{End of C-burning}\\ 
M$_\text{ini}$ & $\upsilon_\text{ini}/\upsilon_\text{crit}$ & $\upsilon_\text{ini}$ & $\bar{v}_\text{MS}$ & $\tau_\text{H}$ & M & $\upsilon_\text{surf}$ & $Y_\text{surf}$ & $\text{N}/\text{C}$ & $\text{N}/\text{O}$ & $\tau_\text{He}$ & M & $\upsilon_\text{surf}$ & $Y_\text{surf}$ & $\text{N}/\text{C}$ & $\text{N}/\text{O}$ & $\tau_\text{C}$ & M & $\upsilon_\text{surf}$ & $Y_\text{surf}$ & $\text{N}/\text{C}$ & $\text{N}/\text{O}$ \\ 
$M_{\sun}$ && \multicolumn{2}{c|}{km s$^{-1}$} & Myr & $M_{\sun}$ & km s$^{-1}$ & \multicolumn{3}{c|}{mass fract.} & Myr & $M_{\sun}$ & km s$^{-1}$ & \multicolumn{3}{c|}{mass fract.} & kyr & $M_{\sun}$ & km s$^{-1}$ & \multicolumn{3}{c|}{mass fract.}\\ 
\hline
$500$ & $0.4$ & $717$ & $646.8$ & $2.1$ & $472.9$ & $104$ & $0.88$ & $51.58$ & $100.04$ & $0.214$ & $462.7$ & $2.5$ & $0.94$ & $83.89$ & $125.68$ & 3.64e-04 & $462.5$ & $0.9$ & $0.94$ & $84.32$ & $125.10$ \\
 & $0$ & -- & -- & $2.0$ & $499.9$ & -- & $0.54$ & $28.65$ & $5.49$ & $0.224$ & $465.9$ & -- & $0.84$ & $58.08$ & $67.32$ & 8.61e-05 & $465.2$ & -- & $0.85$ & $57.17$ & $65.56$ \\
$300$ & $0.4$ & $640$ & $631.2$ & $2.3$ & $286.6$ & $158$ & $0.86$ & $48.97$ & $50.76$ & $0.224$ & $282.4$ & $4.0$ & $0.92$ & $77.22$ & $81.63$ & 4.24e-05 & $282.3$ & $6.4$ & $0.92$ & $78.76$ & $82.13$ \\
 & $0$ & -- & -- & $2.0$ & $300.0$ & -- & $0.25$ & $0.08$ & $0.01$ & $0.230$ & $231.2$ & -- & $0.79$ & $19.27$ & $10.21$ & 2.10e-04 & $221.8$ & -- & $0.78$ & $18.16$ & $1.01$ \\
$200$ & $0.4$ & $584$ & $591.3$ & $2.6$ & $194.1$ & $325$ & $0.49$ & $17.00$ & $2.22$ & $0.242$ & $158.2$ & $0.1$ & $0.82$ & $127.26$ & $25.68$ & -- & -- & -- & -- & -- & -- \\
 & $0$ & -- & -- & $2.2$ & $199.9$ & -- & $0.25$ & $0.08$ & $0.01$ & $0.237$ & $168.2$ & -- & $0.52$ & $44.38$ & $11.16$ & 6.05e-04 & $160.7$ & -- & $0.59$ & $54.54$ & $17.93$ \\
$150$ & $0.4$ & $547$ & $562.2$ & $2.8$ & $146.0$ & $295$ & $0.42$ & $11.44$ & $1.39$ & $0.257$ & $131.5$ & $3.1$ & $0.65$ & $50.96$ & $4.39$ & 6.07e-04 & $127.1$ & $0.0$ & $0.66$ & $54.31$ & $4.63$ \\
 & $0$ & -- & -- & $2.4$ & $149.9$ & -- & $0.25$ & $0.08$ & $0.01$ & $0.249$ & $128.7$ & -- & $0.46$ & $28.66$ & $4.79$ & $0.002$ & $121.1$ & -- & $0.56$ & $45.36$ & $9.46$ \\
120 & $0.4$ & $522$ & $610.0$ & $3.0$ & $118.5$ & $399$ & $0.35$ & $6.94$ & $0.81$ & $0.264$ & $86.6$ & $0.3$ & $0.75$ & $132.69$ & $12.92$ & $0.002$ & $85.0$ & $0.6$ & $0.77$ & $107.09$ & $23.13$ \\ 
 & 0 & -- & -- & $2.5$ & $120.0$ & -- & $0.25$ & $0.08$ & $0.02$ & $0.277$ & $98.4$ & -- & $0.50$ & $40.48$ & $6.00$ & $0.001$ & $96.8$ & -- & $0.53$ & $45.08$ & $7.39$ \\ 
85 & $0.4$ & $483$ & $572.7$ & $3.4$ & $84.3$ & $476$ & $0.31$ & $6.05$ & $0.59$ & $0.300$ & $57.2$ & $0.2$ & $0.75$ & $78.30$ & $67.83$ & $0.003$ & $56.4$ & $0.4$ & $0.79$ & $71.56$ & $55.98$ \\ 
 & 0 & -- & -- & $2.9$ & $85.0$ & -- & $0.25$ & $0.08$ & $0.02$ & $0.284$ & $85.0$ & -- & $0.25$ & $0.08$ & $0.02$ & $0.009$ & $84.4$ & -- & $0.25$ & $0.08$ & $0.02$ \\ 
60 & $0.4$ & $447$ & $503.1$ & $3.9$ & $59.7$ & $547$ & $0.29$ & $8.08$ & $0.55$ & $0.353$ & $56.6$ & $3.2$ & $0.60$ & $18.93$ & $2.00$ & $0.036$ & $56.6$ & $2.3$ & $0.60$ & $22.25$ & $1.88$ \\ 
 & 0 & -- & -- & $3.4$ & $60.0$ & -- & $0.25$ & $0.08$ & $0.02$ & $0.326$ & $60.0$ & -- & $0.25$ & $0.08$ & $0.02$ & $0.016$ & $60.0$ & -- & $0.25$ & $0.08$ & $0.02$ \\ 
40 & $0.4$ & $407$ & $419.0$ & $5.2$ & $39.9$ & $613$ & $0.29$ & $20.29$ & $0.69$ & $0.513$ & $39.8$ & $122$ & $0.43$ & $14.33$ & $5.95$ & $2.090$ & $39.8$ & $165$ & $0.44$ & $14.75$ & $5.92$ \\ 
 & 0 & -- & -- & $4.4$ & $40.0$ & -- & $0.25$ & $0.08$ & $0.02$ & $0.400$ & $40.0$ & -- & $0.25$ & $0.08$ & $0.02$ & $0.071$ & $40.0$ & -- & $0.25$ & $0.08$ & $0.02$ \\ 
30 & $0.4$ & $383$ & $356.5$ & $6.6$ & $30.0$ & $546$ & $0.30$ & $44.43$ & $0.89$ & $0.756$ & $22.9$ & $0.3$ & $0.64$ & $47.64$ & $9.23$ & $0.358$ & $22.0$ & $0.2$ & $0.64$ & $29.69$ & $8.38$ \\ 
 & 0 & -- & -- & $5.5$ & $30.0$ & -- & $0.25$ & $0.08$ & $0.02$ & $0.498$ & $30.0$ & -- & $0.25$ & $0.08$ & $0.02$ & $0.200$ & $30.0$ & -- & $0.25$ & $0.08$ & $0.02$ \\ 
25 & $0.4$ & $367$ & $328.6$ & $7.7$ & $25.0$ & $426$ & $0.30$ & $64.62$ & $1.05$ & $0.611$ & $25.0$ & $137$ & $0.30$ & $66.54$ & $1.07$ & $0.380$ & $25.0$ & $15.5$ & $0.30$ & $68.22$ & $1.08$ \\ 
 & 0 & -- & -- & $6.5$ & $25.0$ & -- & $0.25$ & $0.08$ & $0.02$ & $0.598$ & $25.0$ & -- & $0.25$ & $0.08$ & $0.02$ & $0.419$ & $25.0$ & -- & $0.25$ & $0.08$ & $0.02$ \\ 
20 & $0.4$ & $350$ & $301.6$ & $9.4$ & $20.0$ & $347$ & $0.29$ & $73.03$ & $1.15$ & $0.998$ & $20.0$ & $72.9$ & $0.35$ & $94.20$ & $3.52$ & $1.657$ & $20.0$ & $76.7$ & $0.36$ & $87.59$ & $4.82$ \\ 
 & 0 & -- & -- & $8.0$ & $20.0$ & -- & $0.25$ & $0.08$ & $0.02$ & $0.783$ & $20.0$ & -- & $0.25$ & $0.08$ & $0.02$ & $1.138$ & $20.0$ & -- & $0.25$ & $0.08$ & $0.02$ \\ 
15 & $0.4$ & $330$ & $274.1$ & $13.2$ & $15.0$ & $292$ & $0.28$ & $82.59$ & $1.30$ & $1.353$ & $15.0$ & $53.7$ & $0.33$ & $123.80$ & $2.04$ & $1.754$ & $15.0$ & $1.7$ & $0.40$ & $133.40$ & $29.90$ \\ 
 & 0 & -- & -- & $11.3$ & $15.0$ & -- & $0.25$ & $0.08$ & $0.02$ & $1.171$ & $15.0$ & -- & $0.25$ & $0.08$ & $0.02$ & $3.994$ & $15.0$ & -- & $0.25$ & $0.08$ & $0.02$ \\ 
12 & $0.4$ & $316$ & $257.4$ & $18.0$ & $12.0$ & $263$ & $0.28$ & $79.78$ & $1.35$ & $1.870$ & $12.0$ & $84.3$ & $0.28$ & $82.49$ & $1.37$ & $5.452$ & $12.0$ & $2.0$ & $0.34$ & $141.40$ & $2.46$ \\ 
 & 0 & -- & -- & $15.4$ & $12.0$ & -- & $0.25$ & $0.08$ & $0.02$ & $1.630$ & $12.0$ & -- & $0.25$ & $0.08$ & $0.02$ & $7.104$ & $12.0$ & -- & $0.28$ & $8.37$ & $0.67$ \\ 
9 & $0.4$ & $301$ & $239.4$ & $28.4$ & $9.0$ & $239$ & $0.27$ & $69.56$ & $1.33$ & $2.901$ & $9.0$ & $85.1$ & $0.28$ & $71.65$ & $1.35$ & $4.622$ & $8.9$ & $0.7$ & $0.36$ & $0.03$ & $0.03$ \\ 
 & 0 & -- & -- & $24.2$ & $9.0$ & -- & $0.25$ & $0.08$ & $0.02$ & $2.719$ & $9.0$ & -- & $0.25$ & $0.08$ & $0.02$ & $4.986$ & $9.0$ & -- & $0.26$ & $7.30$ & $0.59$ \\ 
7 & $0.4$ & $286$ & $223.9$ & $43.4$ & $7.0$ & $220$ & $0.27$ & $50.24$ & $1.16$ & $5.20$ & $7.0$ & $7.3$ & $0.27$ & $54.88$ & $1.22$ &&&&&&\\ 
 & 0 & -- & -- & $37.0$ & $7.0$ & -- & $0.25$ & $0.08$ & $0.02$ & $4.896$ & $7.0$ & -- & $0.25$ & $0.08$ & $0.02$ &&&&&&\\ 
5 & $0.4$ & $268$ & $212.0$ & $80.0$ & 5.0 & $205$ & $0.27$ & $34.16$ & $1.00$ & $11.098$ & 5.0 & $63.7$ & $0.27$ & $35.17$ & $1.02$ &&&&&&\\ 
 & 0 & -- & -- & $68.0$ & $5.0$ & -- & $0.25$ & $0.08$ & $0.02$ & $11.101$ & $5.0$ & -- & $0.25$ & $0.08$ & $0.02$ &&&&&&\\ 
4 & $0.4$ & $257$ & $201.7$ & $123.7$ & 4.0 & $192$ & $0.27$ & $23.33$ & $0.84$ & $20.072$ & $4.0$ & $42.7$ & $0.28$ & $47.92$ & $1.26$ &&&&&&\\ 
 & 0 & -- & -- & $103.9$ & $4.0$ & -- & $0.25$ & $0.08$ & $0.02$ & $20.390$ & $4.0$ & -- & $0.25$ & $0.08$ & $0.02$ &&&&&&\\ 
3 & $0.4$ & $233$ & $194.9$ & $224.9$ & 3.0 & $189$ & $0.27$ & $16.85$ & $0.69$ & $40.273$ & 3.0 & $35.7$ & $0.27$ & $17.87$ & $0.72$ &&&&&&\\ 
 & 0 & -- & -- & $185.2$ & $3.0$ & -- & $0.25$ & $0.08$ & $0.02$ & $46.496$ & $3.0$ & -- & $0.25$ & $0.08$ & $0.02$ &&&&&&\\ 
2.5 & $0.4$ & $218$ & $188.9$ & $343.4$ & 2.5 & $191$ & $0.27$ & $13.07$ & $0.60$ & $65.282$ & 2.5 & $38.1$ & $0.28$ & $14.08$ & $0.63$ &&&&&&\\ 
 & 0 & -- & -- & $281.5$ & $2.5$ & -- & $0.25$ & $0.08$ & $0.02$ & $73.509$ & $2.5$ & --  & $0.25$ & $0.08$ & $0.02$ &&&&&&\\ 
2 & $0.4$ & $200$ & $179.9$ & $698.9$ & 2.0 & $185$ & $0.29$ & $11.42$ & $0.55$ & $62.639$ & $1.98$ & $6.6$ & $0.31$ & $26.93$ & $0.83$ &&&&&&\\ 
 & 0 & -- & -- & $546.7$ & $2.0$ & -- & $0.25$ & $0.08$ & $0.02$ & $102.619$ & $1.99$ & -- & $0.25$ & $0.50$ & $0.07$ &&&&&&\\ 
1.7 & $0.4$ & $187$ & $175.6$ & $1211.6$ & 1.7 & $193$ & $0.30$ & $7.82$ & $0.38$ & $90.669$ & $1.67$ & $3.1$ & $0.33$ & $27.75$ & $0.65$ &&&&&&\\ 
 & 0 & -- & -- & $930.8$ & $1.7$ & -- & $0.25$ & $0.08$ & $0.02$ & $119.867$ & $1.68$ & -- & $0.28$ & $1.55$ & $0.17$ &&&&&&\\ 
\hline
\end{tabular}} 
\label{TabListModels} 
\end{table*} 

 \begin{figure*}[h!]
 \centering
\includegraphics{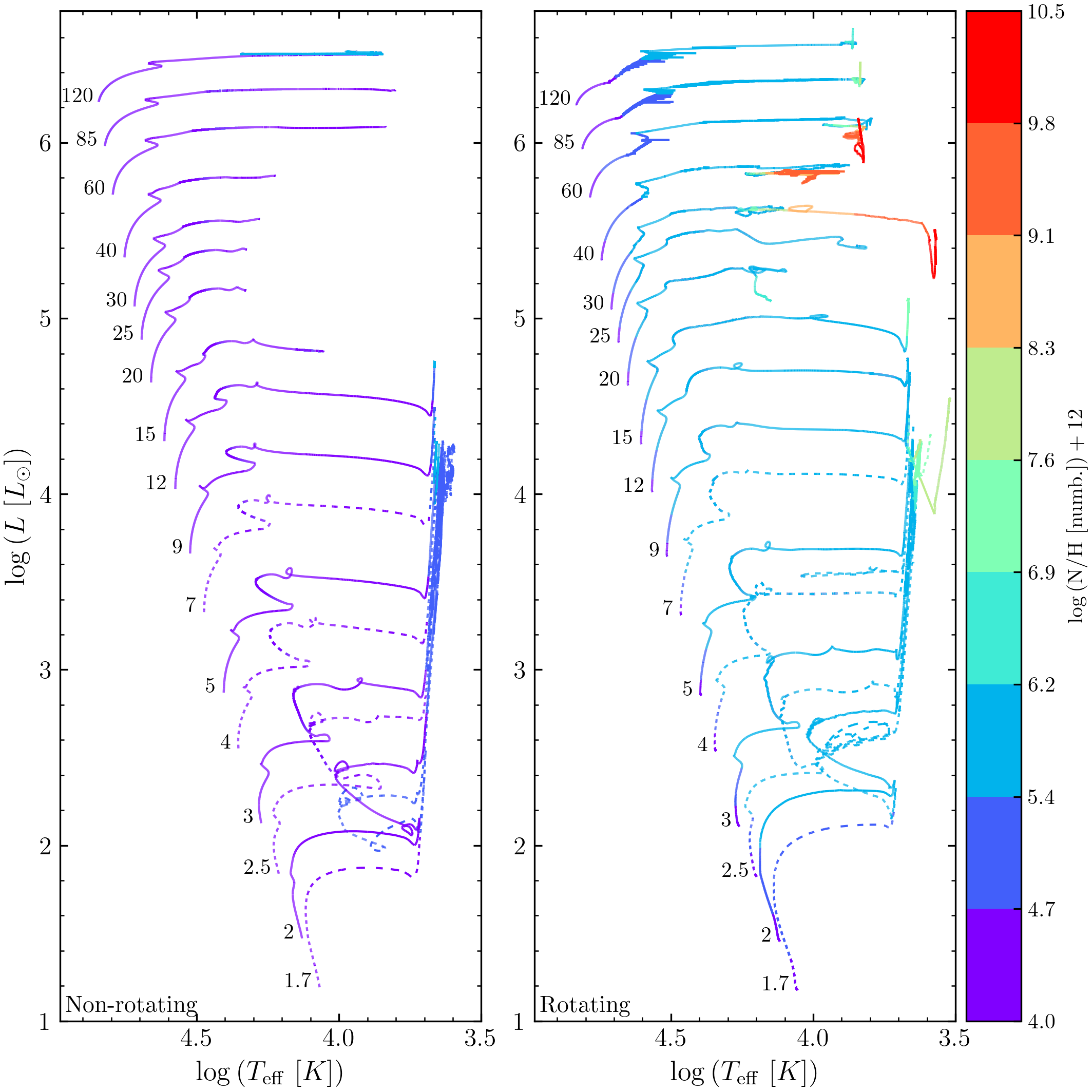}
 \caption{Evolutionary tracks in the HRD for non-rotating (left) and rotating (right) models, colour-coded according to the surface number abundance of nitrogen $\log{(\rm{N/H~[numb.]})} + 12$. Each track is labelled with its initial mass. Every other star below $9\,M_\odot$ is shown with a dashed line in order to better distinguish the models.}
 \label{HRandNH}
 \end{figure*}

Figure~\ref{HRandNH} shows the evolutionary tracks in the Hertzsprung-Russell diagram (HRD) of the present models (up to 120\,$M_\odot$), colour-coded by the surface abundance of nitrogen $\log{(\rm{N/H~[numb.]})} + 12$. We present the main features below:
\begin{itemize}
    \item The width of the MS band increases with the initial mass of the models. This is expected, as it is also the case for the models at other metallicities.
    \item Models (both non-rotating and rotating) below 2.5\,$M_\odot$ undergo complete blue loops during core helium-burning. By complete blue loops we mean here blue loops starting and ending along the Hayashi track. During these, the stars move to higher effective temperatures before going back to the Hayashi track with a higher luminosity. These stars reach the end of the computation in the red part of the HR diagram ($\log{(T_{\rm eff} \rm{[K]})} < 3.7$) before the helium flash or at the early-AGB phase.
    \item Non-rotating models between 2.5 and 12\,$M_\odot$ 
    show partial blue loops in the sense that the blue
    loops start when the stars are still crossing the HR gap. The loops end when the stars evolve redwards again and reach the Hayashi track, where they will stay until the end of the computation. 
    \item Those between 15 and 40\,$M_\odot$ end their lives as blue (we define blue as $\log{(T_{\rm eff}\,\rm{[K]})} > 3.9$) supergiants.
    \item The non-rotating models with initial masses of 60\,$M_\odot$ and above become yellow supergiants (yellow corresponds to effective temperatures between red and blue, $3.7 < \log{(T_{\rm eff}\,\rm{[K]})} < 3.9$), but the 120\,$M_\odot$ model moves back to hotter surface temperatures during core helium-burning as it loses about 20\,\% of its mass.
    \item Rotating models are more luminous during their entire evolution, and generally cooler than their non-rotating counterparts after the MS. They lose more mass (which tends to make hydrogen-poor stars more blue and hydrogen-rich ones more red), and rotation also increases mixing, thus bringing metals to the surface which tends to increase their opacity. This is clearly visible for all models: the surface nitrogen enrichment is much stronger for rotating models than for non-rotating ones. The rotation-induced mixing also brings helium to the surface, which has the effect of decreasing the opacity. In the end, the blue or red fate of stars after the MS is influenced by complex interactions involving convection, rotation, and mass loss. A more thorough study of these effects can be found in \citet{Farrell2020} and \citet{Farrell2022}. We briefly note the redwards extension of the rotating 7\,$M_\odot$ model beyond the Hayashi track. This is linked to the strong mixing undergone by that model. Such a behaviour was already found in \citet[][see their Fig.~10]{MeynetMaeder2002}.
    \item Massive stars can end their lives as red (RSG), yellow (YSG), or blue supergiants (BSG). No single star model ends its life as a stripped Wolf-Rayet (WR) star at this metallicity. Some models finish their evolution with a low surface hydrogen mass fraction ($X_{\rm s}<0.3$), but their effective temperatures are too cold ($\log{(T_{\rm eff}\,\rm{[K]})} < 4$) to be classified as WR stars.
\end{itemize}
 \begin{figure*}[h!]
 \centering
 \includegraphics{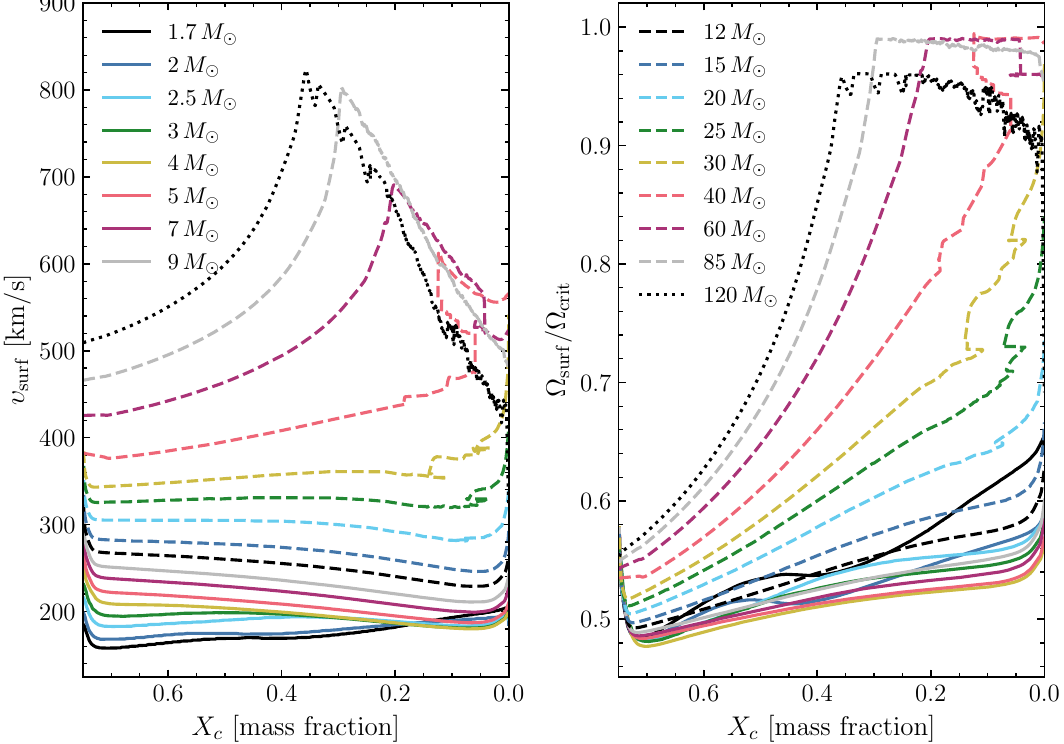}
 \caption{Evolution of the surface rotation velocities of all rotating models during the MS as a function of the central mass fraction of hydrogen $X_{\rm c}$. Left panel: Evolution of the surface equatorial velocities. Right panel: Evolution of the ratio of the angular velocity to the critical angular velocity. All rotating stars computed in this work are shown in each panel, and we split the legend into two. The line styles and colours of both plots represent the same stars in each, whose initial masses are indicated in the legends.}
 \label{Vsurf_OOc_Xc}
 \end{figure*}
 
Figure~\ref{Vsurf_OOc_Xc} shows the evolution of the equatorial velocities $\upsilon_{\rm surf}$ (left panel), and ratios of the angular velocity to the break up (or critical) angular velocity $\Omega_{\rm surf}/\Omega_{\rm crit}$ (right panel) as a function of the central hydrogen mass fraction during the MS. The critical angular velocity corresponds to the rotation rate at which the centrifugal acceleration counteracts gravity at the equator and the surface of the star becomes unbound. Our choice of $\upsilon_{\rm ini}/\upsilon_{\rm crit}=0.4$ corresponds to $\Omega_{\rm ini}/\Omega_{\rm crit}=0.57$ \citep[strictly speaking, slightly varying values of $\Omega_{\rm ini}/\Omega_{\rm crit}$ should be expected for stars of different initial masses, but the variation is very small, see][]{Ekstrom2008}.

The increase in the initial velocity with increasing initial mass is due to our choice of fixed ratio of $\upsilon_{\rm ini}/\upsilon_{\rm crit}=0.4$ on the ZAMS.
For stars with initial masses $M_{\rm ini} \leq 30\,M_\odot$, except for a rapid decrease at the very beginning (barely visible in Fig.~\ref{Vsurf_OOc_Xc}), their equatorial velocities remain roughly constant during most of the MS phase. Due to the contraction when the mass fraction of hydrogen at the centre is only a few percents, a rapid spin-up occurs at the end of the MS. They end the MS at a similar surface velocity as they started. During most of the MS phase, $\Omega_{\rm surf}/\Omega_{\rm crit}$ (right panel) increases. As stars expand during the MS, the critical rotation rate decreases, so the ratio of the equatorial velocity to critical velocity is larger at the end than at the beginning of the MS. The average surface velocity over the MS for these stars (see the fourth column of Table~\ref{TabListModels}) is smaller than the one on the ZAMS.

The models with $M_{\rm ini} \geq 40\,M_\odot$ experience no (or a very slight) spin-down at the beginning of the MS, but their surface velocity increases dramatically compared to less massive models. This spin-up happens earlier, and the maximum velocity reached is larger, the more massive the star. This is a consequence of the fact that meridional currents have larger velocities in those massive stars in the external layers. Let us remind that in those layers the meridional velocity scales as the inverse of the density \citep[see the discussion after Eq. 4.29 in][]{MaederZahn1998}. In these layers angular momentum is brought to the surface by the meridional currents and thus spins up these superficial layers. As a result, these stars may easily reach the critical velocity and thus have part of their mass unbound.
Likely this mass will form a Keplerian disc, and matter as well as angular momentum can be lost: this is what we call the mechanical mass loss. This slows down the surface as shown in the left panel of Fig.~\ref{Vsurf_OOc_Xc}.\\
Once the surface velocity reaches the critical limit during the MS phase, it remains at this limit for the rest of the MS. By removing angular momentum, mass loss makes the surface velocity reach a value below the critical one. On the other hand, meridional currents bring back angular momentum to the surface making the surface velocity reach the critical value again.
Furthermore, as the star expands, the critical rotation velocity decreases, which is why the equatorial velocity also decreases. The plateau in $\Omega_{\rm surf}/\Omega_{\rm crit}$ takes different values between 0.9 and 1 depending on the initial mass of the stars, but this is a numerical and not a physical effect: in order to prevent density inversions and help the convergence of the code for these massive rotating models, a parameter may be introduced to lower the maximum $\Omega_{\rm surf}/\Omega_{\rm crit}$ above which mechanical mass loss happens. This parameter has to be set to lower values for the more massive stars, and we decrease it over the evolution of the 40, 60, 85 and 120\,$M_\odot$ models. For instance for the 120\,$M_\odot$ we had to adjust it as soon as $X_{\rm c}=0.4$, down from 0.99 to 0.96, and then had to decrease it again progressively to 0.90. The mass lost through mechanical mass loss remains modest, as can be seen in the sixth column of Table~\ref{TabListModels}. Another numerical effect causes the vertical oscillations that can be seen on the right panel of Fig.~\ref{Vsurf_OOc_Xc}, as well as the horizontal oscillations near the end of the main sequence on the right panel of Fig.~\ref{HRandNH}; it is due to the way the mechanical mass loss is treated in \textsc{Genec}. For these stars with $M_{\rm ini} \geq 40\,M_\odot$, the average surface velocity over the MS is larger than the one on the ZAMS. Although these stars also expand, the meridional currents that transport angular momentum are strong and compensate for the expansion. More detailed explanations for the behaviour of the surface velocities during the MS can be found in Paper IV \citep{Groh2019}.
\subsection{Evolution of central properties}
\label{Evolution_centre}
 \begin{figure}[h!]
 \centering
 \includegraphics{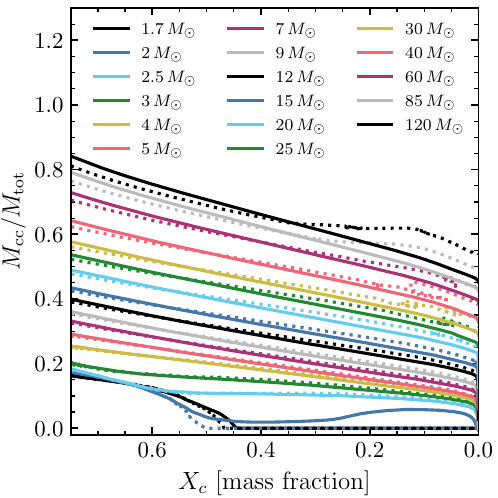}
 \caption{Evolution of the fractional mass of the convective core during the MS as a function of the central hydrogen mass fraction. Solid lines: Non-rotating models. Dotted lines: Rotating models.}
 \label{Mccrel_Xc}
 \end{figure}

Figure~\ref{Mccrel_Xc} shows the evolution of the fractional mass of the convective cores during the MS. Solid lines are for non-rotating models and dotted lines for rotating ones.
On the ZAMS, core masses are smaller for rotating models. This is due to the effect of the centrifugal force that makes a star of a given initial mass follow the track of a star with lower initial mass (keeping the chemical composition the same). Indeed the centrifugal acceleration provides an additional support against gravity, making the pressure gradient comparable to the one in a star of lower gravity or of lower initial mass. This difference is more pronounced for models with higher initial masses, as their cores are initially less dense and more affected by the centrifugal force.

Overall, as expected, the convective cores recede smoothly during the MS. This effect is slightly more pronounced for non-rotating models so that at some point their core mass becomes smaller than that of their rotating counterparts. There are exceptions to this smooth receding of the convective cores: the 1.7 and 2\,$M_\odot$ models, and the rotating models of 25, 30, and 40\,$M_\odot$. We discuss these two distinct cases below.

In the lowest-mass models, the convective core recedes faster than in the other models. It completely disappears around $X_{\rm c}\sim0.5$, meaning that the core then becomes entirely radiative. As discussed in \citet[Sect.~3.2]{Groh2019}, this results from the interplay between the pp chains and the CNO cycle. At the beginning of evolution, the star is chemically homogeneous and therefore contains some, although very little considering the extremely low metallicity, CNO elements. These fuel the CNO cycle, generating more energy (and thus a larger temperature gradient) than the pp chains and keep the core convective. At a certain point (corresponding to $X_{\rm c}\sim0.5$), the energy generation drops and the core becomes radiative. Carbon and oxygen are very much depleted (and have been converted into nitrogen). However, they are also depleted in similar proportions in higher-mass models, yet the CNO cycle continues. The reason is that these have hotter cores. As a result, they can maintain a large radiative gradient (greater than the adiabatic gradient), thus keeping at least a small part of their core convective. It is interesting to note \citep[as had already been noted in][]{Georgy2013} that a very slight change in central temperature will make a star pass from the CNO cycle to the pp-chains.\\
In the rotating models, two counteracting effects are at play: on the one hand, rotational mixing brings carbon and oxygen from the envelope into the core; on the other hand, rotation supports part of the weight of the star through the centrifugal force. The former increases the `fuel' supply of the CNO cycle while the latter decreases the temperature of the core (this decrease in temperature makes the energy generation by the CNO cycle much smaller, due to its very strong dependence on temperature). We can see on Fig.~\ref{Mccrel_Xc} that the second effect is the stronger one, as the core becomes radiative earlier in the rotating 1.7\,$M_\odot$ star than in the non-rotating one; and it becomes radiative in the rotating 2\,$M_\odot$ model whereas it remains convective in the non-rotating one.

The 25, 30, and 40\,$M_\odot$ rotating models experience breathing pulses (which can also be seen in Fig.~\ref{Vsurf_OOc_Xc}) towards the end of the MS: around $X_{\rm c}\sim0.2-0.1$ hydrogen is brought into the core from the envelope, increasing $X_{\rm c}$ (the curves move back to the left) as well as the size of the convective core. It is interesting to note that breathing pulses usually appear in stellar evolution models during core helium-burning, and they bring newly formed helium into the core from the hydrogen-burning shell; in the present case we see them also during the MS.\\
For the models with $M_{\rm ini}\geq 60\,M_\odot$, we do not see breathing pulses. The effect of rotation is simply to increase the mass of the convective core relative to the non-rotating models, from $X_{\rm c}\sim0.3$ onwards.\\

\begin{figure*}[h!]
\centering
\includegraphics{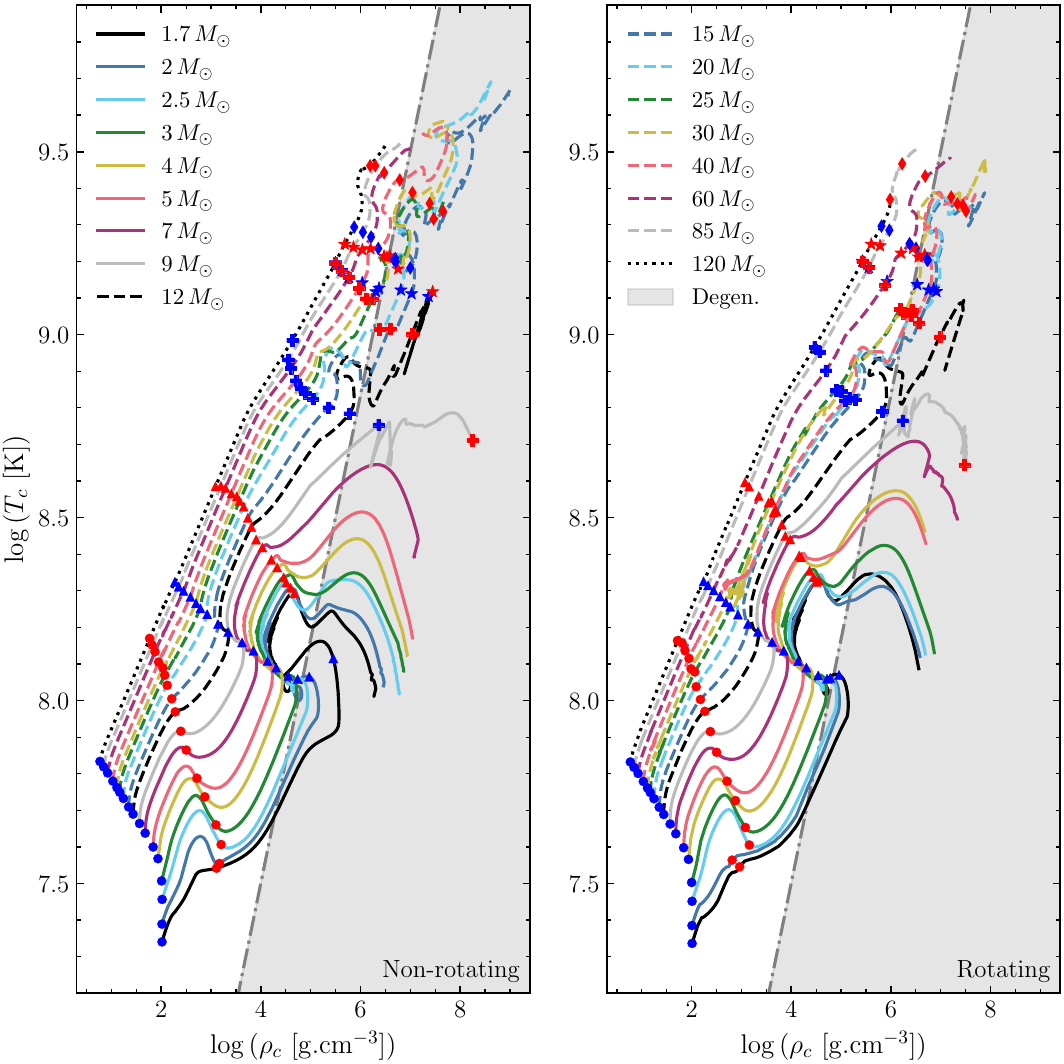}
\caption{Evolution of the central densities and temperatures for all the models computed, starting at the ZAMS. Left panel: Non-rotating models. Right panel: Rotating models. The line styles and colours of both plots represent the same stars in each, whose initial masses are indicated in the legends. The blue (red) markers indicate the beginning (end) of each burning phase (circles: H, triangles: He, crosses: C, stars: Ne, diamonds: O). The grey-shaded areas correspond to the degenerate region.}
\label{rhoT}
\end{figure*}
Figure~\ref{rhoT} shows the evolution of the central densities and temperatures of the present models, starting at the ZAMS. The blue (red) markers indicate the beginning (end) of each burning phase (circles: H, triangles: He, crosses: C, stars: Ne, diamonds: O).

On the ZAMS, as expected, central temperature (density) is an increasing (decreasing) function of stellar mass. All models' cores contract during the MS. There is a hook at the end of the MS for the lower masses of the grid (up to 15\,$M_\odot$), except for both the 1.7\,$M_\odot$, and the rotating 2\,$M_\odot$ models. This echoes the behaviour of the convective core mass discussed above. Indeed, since the CNO cycle stops during the MS in these three stars, the core temperature does not drop as much, when central hydrogen-burning ends, as in the higher-mass stars where the CNO cycle has been heating the core. For the stars between 2.5 and 15\,$M_\odot$, the peak in central temperature before the end of the MS happens when the mass fraction of hydrogen in the core is around 2\,\%. At that stage the core becomes more and more radiative. This implies that
on the whole it expands (convective regions are more compact). The core is likely very near isothermality, so a larger fraction of the pressure gradient comes from the density gradient, hence the increase in the central density.

\subsection{Very massive stars} \label{VMS}
\begin{figure*}
    \centering
    \includegraphics{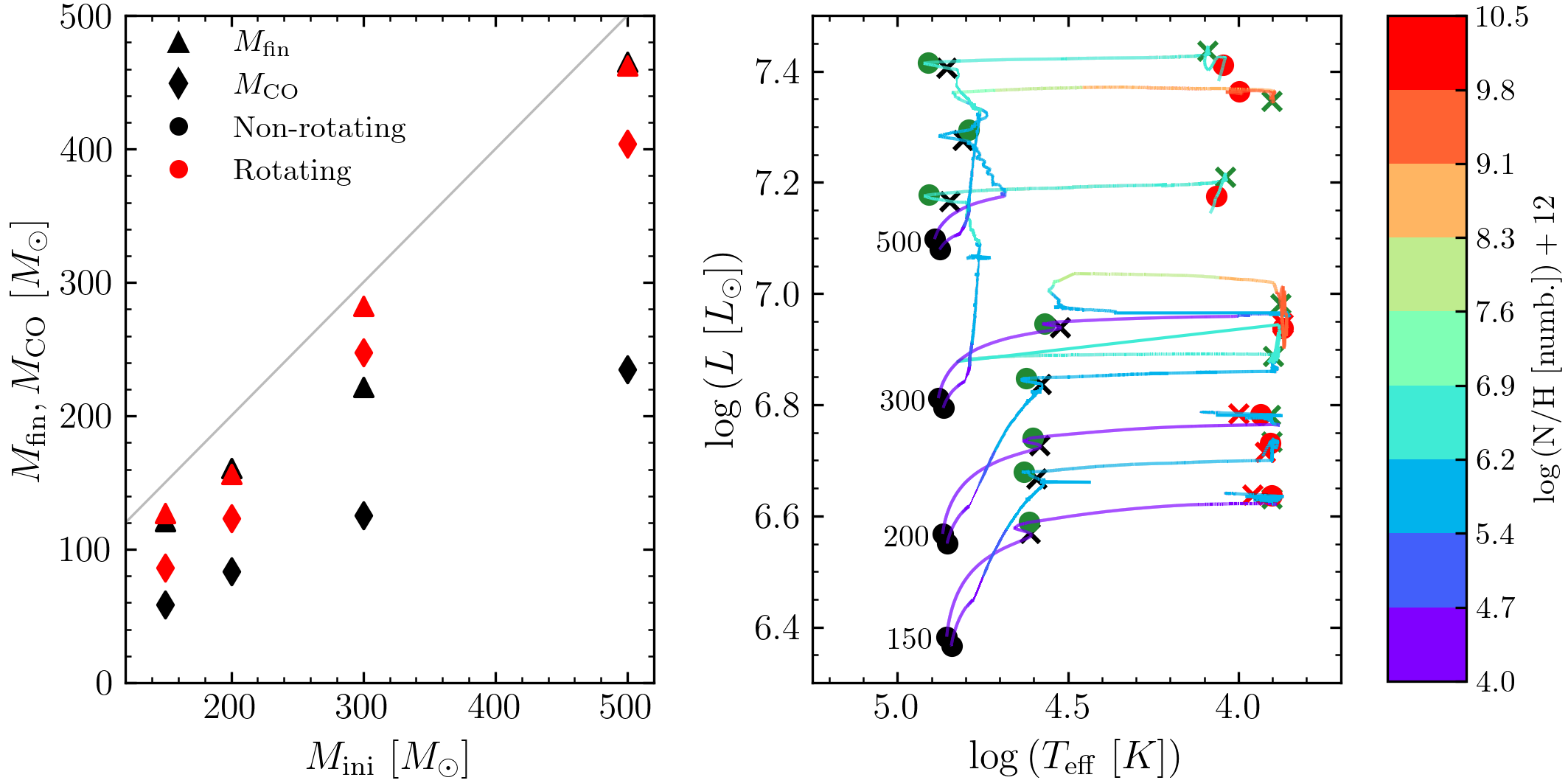}
    \caption{Characteristic masses and evolution in the HRD for the VMSs. Left panel: Final mass (triangles) and mass of the CO core (diamonds) at the end of core helium-burning for non-rotating (black) and rotating (red) VMS models. The grey line corresponds to $y=x$. Right panel: Evolutionary tracks in the HRD, colour-coded according to the surface enrichment of nitrogen $\log{({\rm N} / {\rm H~[numb.]})}+12$. Each pair of (non-rotating, rotating) tracks is labelled with the initial mass of the models. The circles and crosses respectively indicate the beginning and end of each central burning phase (black: hydrogen, green: helium, red: carbon). The non-rotating tracks are the ones that remain purple during the MS.}
    \label{fig:HRD_VMS}
\end{figure*}
Here we present the properties of the VMS models we computed, with initial masses of 150, 200, 300, and 500\,$M_\odot$. We again computed two sets, one of non-rotating models and one of rotating models with $\upsilon_{\rm ini}/\upsilon_{\rm crit}=0.4$. Figure~\ref{fig:HRD_VMS} shows characteristic masses (left panel) and the evolution in the HRD (right panel) for these models. On the left panel we show the final mass and the mass of the CO core at the end of core helium-burning. On the right panel, the tracks are colour-coded by the surface abundance of nitrogen. The non-rotating tracks are the ones that remain purple during the MS. The rotating 200\,$M_\odot$ model reaches the end of core helium-burning, and all the other models reach at least the end of the core carbon-burning phase.

Interestingly, rotating models do not end their lives with substantially lower masses than non-rotating ones. As a matter of fact, the 150 and 300\,$M_\odot$ rotating models have larger final masses than their non-rotating counterparts. The mass-loss history of these models depends on their rotation: non-rotating models reach the end of the MS having lost a minimal ($<0.2\,M_\odot$) amount of mass while rotating ones lose between 4 and 27\,$M_\odot$ during the MS (increasing as a function of initial mass). Conversely, during core helium burning the non-rotating models lose between 21 and 69\,$M_\odot$, whereas the rotating ones lose between 4\,$M_\odot$ and 36\,$M_\odot$. So even though the rotating models lose more mass during the main sequence, they finish the core helium-burning phase at a similar total mass (except for the 300\,$M_\odot$ stars where the non-rotating model loses a lot more mass). These large differences in mass-loss history are due to the different evolutionary paths of the stars. During the main sequence, rotating stars have higher luminosities and lower surface gravities than their non-rotating counterparts, which makes them lose more mass. During the core helium-burning and carbon-burning phases, the non-rotating stars are cooler than the rotating ones. They spend more time with $\log{(T_{\rm eff}\,[K])}<3.9$, which leads to stronger mass losses. This is especially noticeable for the 300\,$M_\odot$ models, and indeed these models boast the largest difference in their final masses.\\
We note that the Eddington factor $\Gamma_{\rm e}$ for these stars reaches values of $\Gamma_{\rm e}=0.7-0.9$ at the end of the main sequence. As mentioned in Sect.~\ref{Ingredients}, we do not use a specific mass-loss rate for stars with high $\Gamma_{\rm e}$. As a result we may underestimate the mass lost by the VMSs.

Rotating VMS models have more massive convective cores, for instance in the 300\,$M_\odot$ rotating model the convective core makes up 89\,\% of the total mass of the star at the end of the MS, against 50\,\% in the non-rotating one. This difference increases with larger initial mass, and it  follows the trend that can be seen for the massive stars 60\,$M_\odot$ and above. For these stars, the convective core does not recede as much in the rotating case than in the non-rotating one towards the end of the MS (see the discussion in Sect.~\ref{Evolution_centre} and Fig.~\ref{Mccrel_Xc}). In the rotating models, the convective cores remain roughly constant in mass starting from $X_{\rm c}\sim0.4-0.5$ already (earlier for more massive stars). This point where the convective core stops decreasing corresponds to the angular point in the HRD, which can be seen for the 60, 85, and 120\,$M_\odot$ models in Fig.~\ref{HRandNH}, and in Fig.~\ref{fig:HRD_VMS} for the 150, 200, 300, and 500\,$M_\odot$ ones. We note that this angular point in the HRD track also appears for the non-rotating 500\,$M_\odot$ model. Since most of the mass of these stars is part of their convective core, they are almost chemically homogeneous. This explains why their evolution during the MS follows a more vertical path than the redwards evolution of their non-rotating counterparts (a fully chemically homogeneous evolution would lead to a bluewards evolution, which is not the case here). In the case of the non-rotating 500\,$M_\odot$ star, interactions between the convective core and convective regions above the core lead to significant enrichment of the outer layers,
and as a result its track evolves bluewards.\\
The CO core masses are larger for the rotating models than for non-rotating ones, which is a consequence of the more massive convective cores during the MS: a more massive convective core will have access to more material to fuse. As a result the mass coordinate within the star below which the mass fraction of helium $Y<10^{-2}$ will be larger.

The mass lost by these VMSs is enriched in hydrogen-burning products ($^{4}$He and $^{14}$N), and it is depleted in hydrogen, carbon, and oxygen. Indeed, while the initial mass fraction of $^{14}$N in these stars is $x(^{14}{\rm N})_{\rm ini}=1.02\times10^{-7}$, in their winds it is $x(^{14}{\rm N})_{\rm winds}\sim4-5\times10^{-6}$ (apart from the non-rotating 300 and 500\,$M_\odot$, with $x(^{14}{\rm N})_{\rm winds}\sim2\times10^{-3}~\text{and}~5\times10^{-3}$ respectively, individual values vary only slightly among the other six VMS models regardless of whether they are rotating or not). This nitrogen is produced in the hydrogen-burning core during the MS (and the hydrogen-burning shell during more advanced phases of evolution) via the CNO cycle, and subsequently moves to the surface via convection (for non-rotating stars) or rotation-induced mixing. Interestingly, we find no significant difference between non-rotating and rotating models of 150 and 200\,$M_\odot$ in how nitrogen-rich their winds are. In other words, similar amounts of $^{14}$N are produced in H-burning regions, and the two mixing processes have the same efficiency in bringing this nitrogen to the surface. We can see on Fig.~\ref{fig:HRD_VMS} that the surface abundance of nitrogen of the non-rotating models only starts increasing when they reach $\log{(T_{\rm eff}~[K])}\sim 4$. The 300 and 500\,$M_\odot$ non-rotating models see their surface abundance of nitrogen reach $\log{({\rm N/H~[numb.]})}+12\sim9.6$, compared to $\log{({\rm N/H~[numb.]})}+12\sim6-7$ for all the other models. This 3\,dex difference is the reason behind the 3 order-of-magnitude difference in $x(^{14}{\rm N})_{\rm winds}$ between these two stars and the rest.

\section{Effects of metallicity}
\label{diffZ}
This paper constitutes the last of the \textsc{Genec} grids computed with the same set of input physics as \citet{Z0142012}. Therefore, the isolated effect of metallicity on single star evolution can be investigated. In this section we discuss the impact of metallicity on the evolution in the HRD, the internal properties of the stars, their fates, and $^{14}$N production. A detailed discussion of the fates of single stars, and their light element yields, will be presented in Hirschi et al. (in prep.).

\subsection{Surface properties}
\label{diffZ_surface}
\begin{figure*}[h!]
\centering
\includegraphics{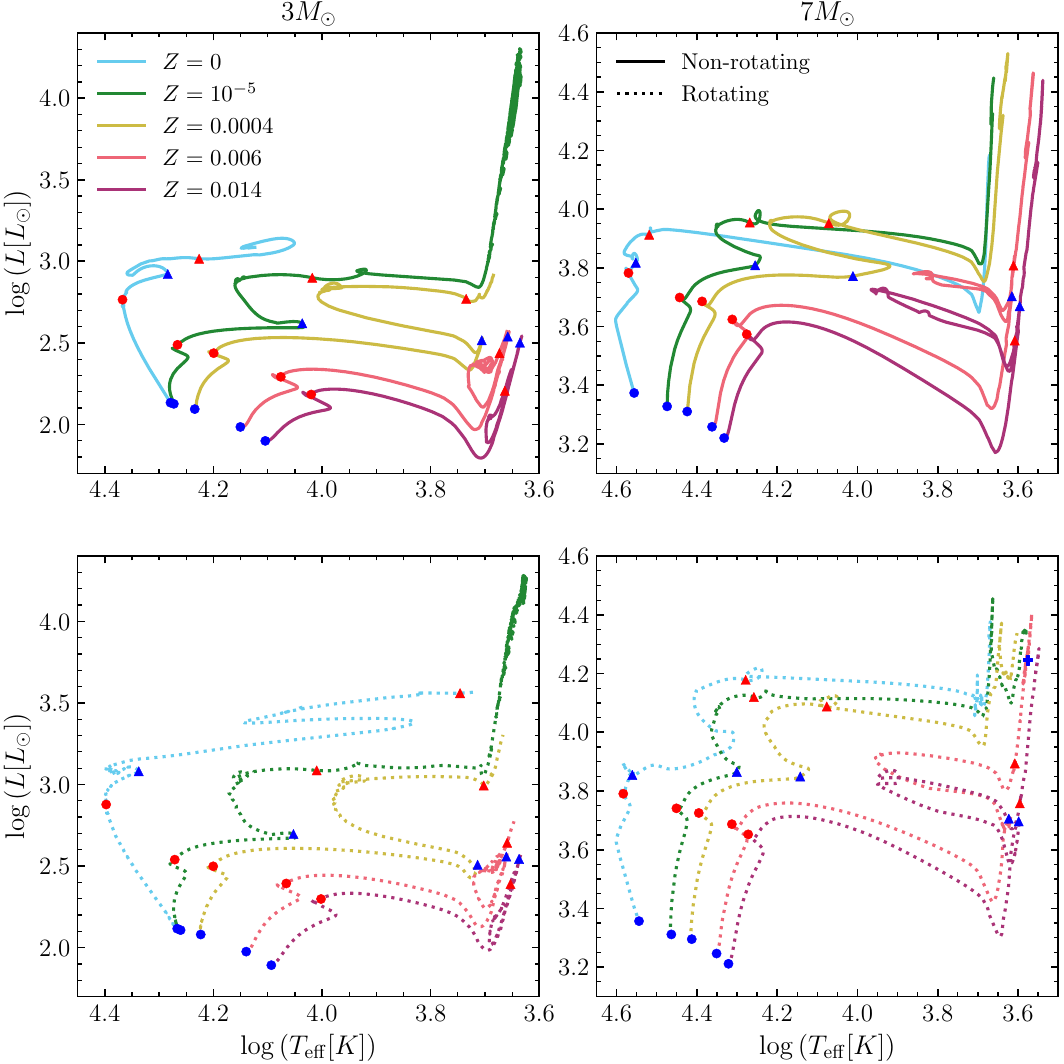}
\caption{Evolutionary tracks in the HRD of the 3\,$M_\odot$ (left column) and 7\,$M_\odot$ (right column) models at five different metallicities ($Z=0$ \citep[Pop~III, from][blue]{Pop32021}, $Z=10^{-5}$ (EMP, this paper, green), $Z=0.0004$ \citep[I Zw 18, from][yellow]{Groh2019}, $Z=0.006$ \citep[Large Magellanic Cloud (LMC), from][pink]{Z0062021}, and $Z=0.014$ \citep[solar, from][purple]{Z0142012}). Top panel: Non-rotating models (solid lines). Bottom panel: Rotating models (dotted lines).}
\label{HRD_diffZ}
\end{figure*}

\begin{figure}[h!]
\centering
\includegraphics{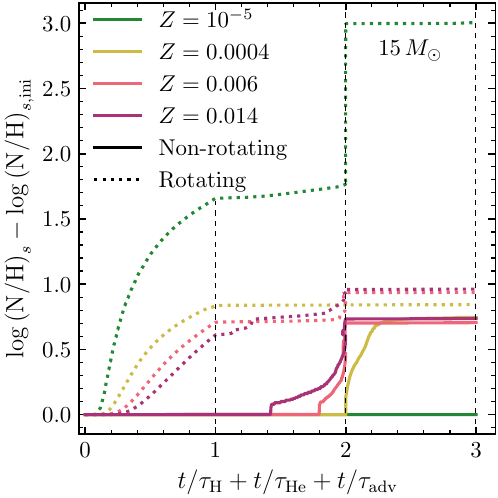}
\caption{$^{14}$N surface enrichment $\log{(\text{N}/\text{H})}_{\rm s}-\log{(\text{N}/\text{H})}_{\rm s, ini}$ over the evolution of the 15\,$M_\odot$ models at different metallicities ($Z=10^{-5}$ (this paper, green), $Z=0.0004$ (I Zw 18, yellow), $Z=0.006$ (LMC, pink), $Z=0.014$ (solar, purple)). The horizontal axis goes from 0 to 1 for the MS, from 1 to 2 for core helium burning, and from 2 to 3 for the rest of the advanced phases. The $Z=0$ models are not shown because their initial $^{14}$N is 0. Solid lines: Non-rotating models. Dotted lines: Rotating models.}
\label{NHs_t_trel_diffZ_15M}
\end{figure}

Figure~\ref{HRD_diffZ} shows the evolution in the HRD of the 3 and 7\,$M_\odot$ models computed at different metallicities ($Z=0$ \citep[Pop~III, from][blue]{Pop32021}, $Z=10^{-5}$ (EMP, this paper, green), $Z=0.0004$ \citep[I Zw 18, from][yellow]{Groh2019}, $Z=0.006$ \citep[LMC, from][pink]{Z0062021}, and $Z=0.014$ \citep[solar, from][purple]{Z0142012}).

For all masses during the MS (and in general throughout the entire stellar evolution), stars are more luminous and blue the lower their metallicity. This is expected since metals increase opacity, so that increasing the metallicity makes a star's envelope absorb more photons. This means that the energy of these photons will be transferred to the envelope, making it expand and be cooler at the surface. Also, fewer photons leave through the surface of the star, decreasing its luminosity. We notice the same effect for stars on the Hayashi track, where those at higher metallicity have a lower effective temperature and luminosity than their lower-metallicity counterparts.

The left hook at the end of the MS is more pronounced for higher metallicity models. Especially noticeable is the absence of hook for the 3\,$M_\odot$ Pop~III models (they finish the MS at $\log{(T_{\rm eff}~[K])}\sim 4.3-4.4$ and $\log{(L~[L_\odot])}\sim2.7$, the bluewards evolution which looks like a hook happens after core helium ignition). This hook is caused by the contraction of the whole star at the end of the MS (heating up the core to maintain the energy production). Since the temperature of the core at the end of the MS decreases with increasing metallicity, the models at higher metallicity need to contract more than those at lower metallicity, thus making a more pronounced hook.

Similarly, during the subgiant phase, stars at higher metallicity cross the entirety of the HRD until they start igniting helium in the core at a lower effective temperature (e.g. $T_{\rm eff}\sim3.6-3.7$ for the 3 and 7\,$M_\odot$ models at LMC and solar metallicities, against $T_{\rm eff}\sim4-4.2$ for the same models at $Z=10^{-5}$). This is because core helium-burning starts at a similar temperature ($\log{(T_{\rm c}~{\rm [K]})}\sim8.1-8.2$) for all stars no matter their mass or metallicity. Since the stars at higher metallicity have lower central temperatures at the end of the MS, their cores needs to contract more in order to reach the required temperature. This causes the envelopes of these stars at higher metallicity to expand more, thus starting core helium-burning as redder stars than their lower metallicity counterparts.

The effect of metallicity on blue loops can be seen clearly for the 7\,$M_\odot$ stars, where the blue loops are more extended for stars at higher metallicities. At $Z=0.0004$, we see that effective temperature increases from $\log{(T_{\rm eff}~{\rm [K]})}=3.9$ to $\log{(T_{\rm eff}~{\rm [K]})}=4.2$ then decreases to $\log{(T_{\rm eff}~{\rm [K]})}=4$ at the end of core helium-burning, but this happens over the entirety of that phase instead of only towards its end. Similarly for $Z=10^{-5}$ and $Z=0$, the effective temperature first decreases sharply before slowly increasing during core helium-burning, and finally dropping as helium is depleted in the core; this effect is less pronounced the lower the metallicity.

Regarding the 60\,$M_\odot$ models, the ones at LMC and solar metallicities become WR stars during core helium-burning, and they end their evolution with the largest effective temperatures ($\log{(T_{\rm eff}~{\rm [K]})}\sim5-5.4$). The models at $Z=0.0004$, $Z=10^{-5}$, and $Z=0$ do not become WR stars and their final effective temperatures increase with decreasing metallicity (except for the rotating $Z=10^{-5}$ model which has the lowest final effective temperature of all the 60\,$M_\odot$ models). In the context of single star evolution, WR stars are the result of strong mass loss. Since mass loss is positively correlated with metallicity via the effect of opacity, it makes sense that the WR phase is reached for high metallicity models but not for lower ones.

Rotation unequivocally increases the luminosity of models of all masses and metallicities. However, its effect on effective temperatures is less straightforward. This is the result of the interplay between chemical mixing and angular momentum transport processes, with mass and metallicity. It is complicated to predict the effect of rotation as its impact will depend on the mass and metallicity of a model, as well as the actual initial rotation speed (that is why we perform numerical computations). We remind the reader that the effects shown by our models are contingent on our choice of input physics, as well as the moderate ($\upsilon_{\rm ini}/\upsilon_{\rm crit}=0.4$) initial rotation velocity of the stars.\\

Figure~\ref{NHs_t_trel_diffZ_15M} shows the evolution of the relative surface nitrogen enrichment $\log{(\text{N}/\text{H})}_{\rm s}-\log{(\text{N}/\text{H})}_{\rm s, ini}$ for the 15\,$M_\odot$ models at different metallicities ($Z=10^{-5}$ (this paper, green), $Z=0.0004$ (I Zw 18, yellow), $Z=0.006$ (LMC, pink), $Z=0.014$ (solar, purple)). The horizontal axis goes from 0 to 1 for the main sequence, from 1 to 2 for core helium-burning, and from 2 to 3 for the rest of advanced phases. The $Z=0$ models are not shown because their initial $^{14}$N is 0 and as a result one cannot define a relative nitrogen enrichment for these stars. Solid lines show non-rotating models, and dotted lines rotating ones. Surface $^{14}$N enrichment can be considered a proxy for the efficiency of mixing \citep{Dufton2020,Dufton2024,Bouret2021,Wessmayer2022,Aschenbrenner2023}.

For all metallicities, rotation increases the strength of surface $^{14}$N enrichment. This is unsurprising and is caused by rotational mixing. In these rotating models, mixing is stronger at lower metallicities, especially during the main sequence. The reason for stronger mixing with lower metallicity is that the mixing timescale positively correlates with $r^2/D$, where $r$ is the radius and $D$ the diffusion coefficient. Stars at lower metallicity are more compact (smaller $r$), making the mixing timescale smaller, thus increasing the efficiency of mixing.

In the post-MS phases of evolution however, another phenomenon can greatly affect the surface $^{14}$N abundances: the convective envelope. Indeed, when a star forms a convective envelope, it will mix that entire zone on an almost instantaneous timescale (compared to its evolutionary timescale). A convective envelope forms when the temperature gradient in the envelope is large, which means that it forms more readily when the star's surface is cooler and the opacities in the envelope are larger. Looking at their tracks in the HRD (not shown graphically), the non-rotating $Z=0$ and $Z=0.0004$ 15\,$M_\odot$ models burn helium and carbon in their cores at much larger effective temperatures than their counterparts at other metallicities. As a result, they do not form convective envelopes. This keeps their surface $^{14}$N abundances constant during the post-MS phases of evolution. Conversely, the rotating 15\,$M_\odot$ model at $Z=10^{-5}$ burns carbon in its core as a RSG with $\log{(T_{\rm eff}~{\rm [K]})}\sim3.65$, develops an extended convective envelope, and sees an almost instantaneous 50-fold increase in its surface $^{14}$N abundance at the onset of carbon-burning in the core.

In non-rotating models, enrichment occurs earlier in stellar evolution for higher metallicities. This is also a consequence of the convective envelope, which appears earlier at higher metallicity because the stars become red supergiants earlier. The non-rotating $Z=10^{-5}$ model always remains blue, and as such sees no surface $^{14}$N enrichment.

\subsection{Internal properties and lifetimes}
\label{diffZ_internal}
 \begin{figure*}[h!]
 \centering
\includegraphics{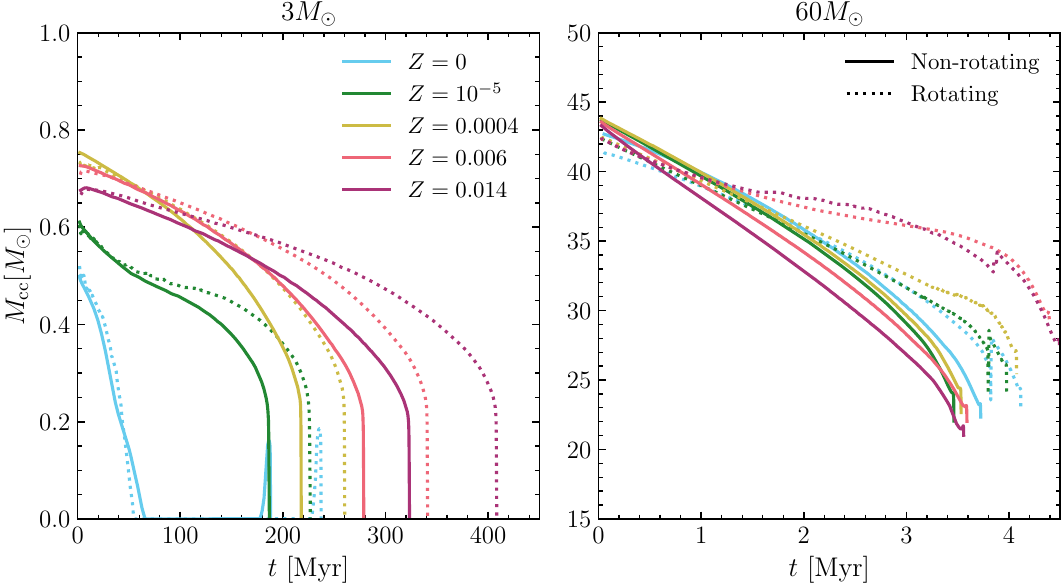}
 \caption{Evolution of the convective core mass during the main sequence for the 3\,$M_\odot$ (left panel) and 60\,$M_\odot$ (right panel) models at different metallicities ($Z=0$ (Pop~III, blue), $Z=10^{-5}$ (this paper, green), $Z=0.0004$ (I Zw 18, yellow), $Z=0.006$ (LMC, pink), and $Z=0.014$ (solar, purple)). Solid lines: Non-rotating models. Dotted lines: Rotating models.}
 \label{Mcc_diffZ}
 \end{figure*}
 \begin{figure*}[h!]
 \centering
 \includegraphics{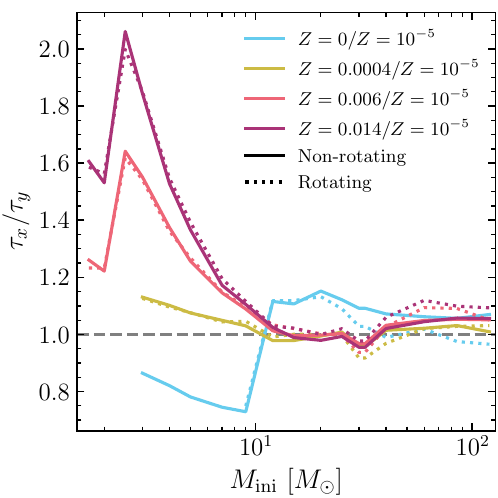}\hspace{33pt}\includegraphics{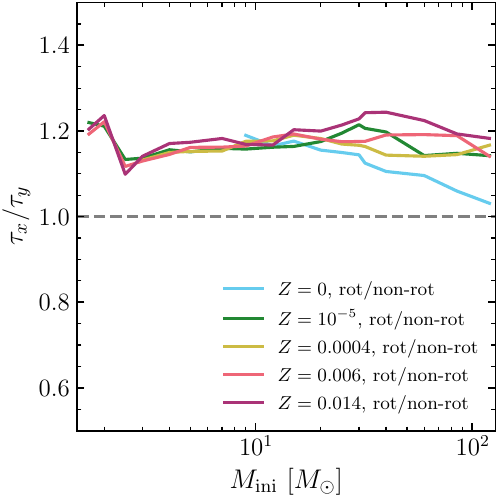}
 \caption{Ratios between  lifetimes of different sets of models. Left panel: Ratios between the lifetimes of models at different metallicities ($Z=0$ (Pop~III, blue), $Z=0.0004$ (I Zw 18, yellow), $Z=0.006$ (LMC, pink), and $Z=0.014$ (solar, purple)), and those of models computed in this paper ($Z=10^{-5}$) as a function of the initial stellar mass $M_{\rm ini}$. Solid lines: Non-rotating models. Dotted lines: Rotating models. Right panel: Ratios between the lifetimes of rotating and non-rotating models at all metallicities ($Z=0$ (Pop~III, blue), $Z=10^{-5}$ (this paper, green), $Z=0.0004$ (I Zw 18, yellow), $Z=0.006$ (LMC, pink), $Z=0.014$ (solar, purple)). In both panels, the horizontal axes have a logarithmic scale.}
 \label{lifetimes_diffZ}
 \end{figure*}

Figure~\ref{Mcc_diffZ} shows the evolution of the masses of the convective cores for the 3\,$M_\odot$ (left column) and 60\,$M_\odot$ (right column) models at different metallicities ($Z=0$ (Pop~III, blue), $Z=10^{-5}$ (this paper, green), $Z=0.0004$ (I Zw 18, yellow), $Z=0.006$ (LMC, pink), $Z=0.014$ (solar, purple)). Solid lines show non-rotating models and dotted lines, rotating ones.

Apart from the $Z=0$ models, rotation increases the mass of the convective core for all metallicities, and this increase is more pronounced in stars with higher metallicity and mass.\\
For the 60\,$M_\odot$ models, even though the absolute mass of the convective core decreases with metallicity (because mass loss processes are stronger), its relative mass at the same point in the MS (characterised by the same value of $X_{\rm c}$) is an increasing function of the metallicity. This is not strictly the case for the 3\,$M_\odot$ models, where the $Z=0.006$ and $Z=0.0004$ models have slightly larger convective cores than the $Z=0.014$ model (when comparing them at the same value of $X_{\rm c}$).\\

Figure~\ref{lifetimes_diffZ} shows the ratios between the lifetimes of different sets of models. The left panel shows the ratios between the lifetimes models at different metallicities ($Z=0$ (Pop~III, blue), $Z=0.0004$ (I Zw 18, yellow), $Z=0.006$ (LMC, pink), $Z=0.014$ (solar, purple)), and those of current models, as a function of the initial stellar mass. Solid lines show non-rotating models and dotted lines rotating ones. A value higher than 1 means that the model at the corresponding metallicity has a longer lifetime than the one at $Z=10^{-5}$. The right panel shows the ratio between the lifetimes of rotating and those of non-rotating models at the same metallicities (same colours), as well as for the models of the current paper ($Z=10^{-5}$, green). The horizontal axes of both panels are in logarithmic scale.

For masses below $10-20\,M_\odot$ (depending on metallicity), longer lifetimes are obtained for models of higher metallicity. The lifetime ratios between the $Z=10^{-5}$ and the other models do not depend on rotation. In other words, the ratio between the lifetimes of rotating and non-rotating models does not depend on metallicity in this mass range (rotating models have 10-20\,\% longer lifetimes depending on the mass, see the right panel of Fig.~\ref{lifetimes_diffZ}).

For masses above 20\,$M_\odot$, there is no clear trend, especially between 20 and 40\,$M_\odot$. Lifetimes seem to increase with metallicity for non-rotating models, but this result is blurred by rotation. In this mass range the largest difference is between zero-metallicity models and the rest. We see that the ratio between the lifetimes of rotating and non-rotating models decreases with increasing mass for these Pop~III models, whereas it remains roughly constant for all the other metallicities.\\
Above 40\,$M_\odot$, the $Z=10^{-5}$ models are the shortest-lived of all metallicities (except for the rotating Pop~III models above 85\,$M_\odot$). In any case, the lifetime differences in this $M_{\rm ini} > 20 M_\odot$ mass range are less than 10\,\%, which is modest compared to the ones in the lower mass range (up to twice longer for the $Z=0.014$ models with $M_{\rm ini}=2.5\,M_\odot$).

\subsection{Final properties and nucleosynthesis}
\label{diffZ_final}
In this section, we present the final mass of the models at different metallicities, as well as their fate (what kind of remnant they will leave after their death). We also discuss the stellar yields of $^{14}$N. A full exploration of all the elemental yields for all the \textsc{Genec} models at all metallicities is beyond the scope of this paper and is deferred to a future standalone study.\\

 \begin{figure}[h!]
 \centering
 \includegraphics{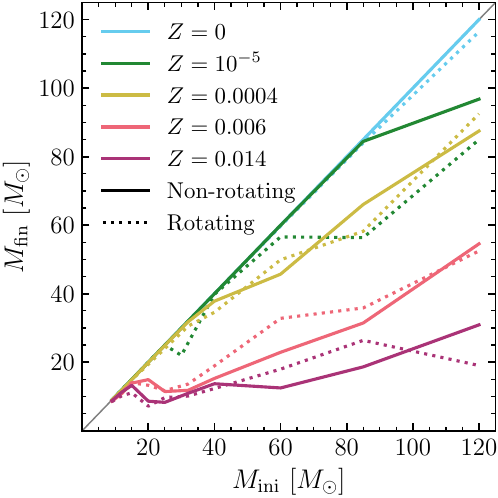}
 \caption{Final mass $M_{\rm fin}$ (mass of the last computed model) as a function of the initial mass $M_{\rm ini}$ for all models with initial mass $M_{\rm ini}\geq9M_\odot$ at different metallicities ($Z=0$ (Pop~III, blue), $Z=10^{-5}$ (this paper, green), $Z=0.0004$ (I Zw 18, yellow), $Z=0.006$ (LMC, pink), $Z=0.014$ (solar, purple)). Solid lines: Non-rotating models. Dotted lines: Rotating models. The thin grey line corresponds to $M_{\rm fin}=M_{\rm ini}$.}
 \label{Mfin_Mini_diffZ}
 \end{figure}
 \begin{figure*}[h!]
 \centering
 \includegraphics{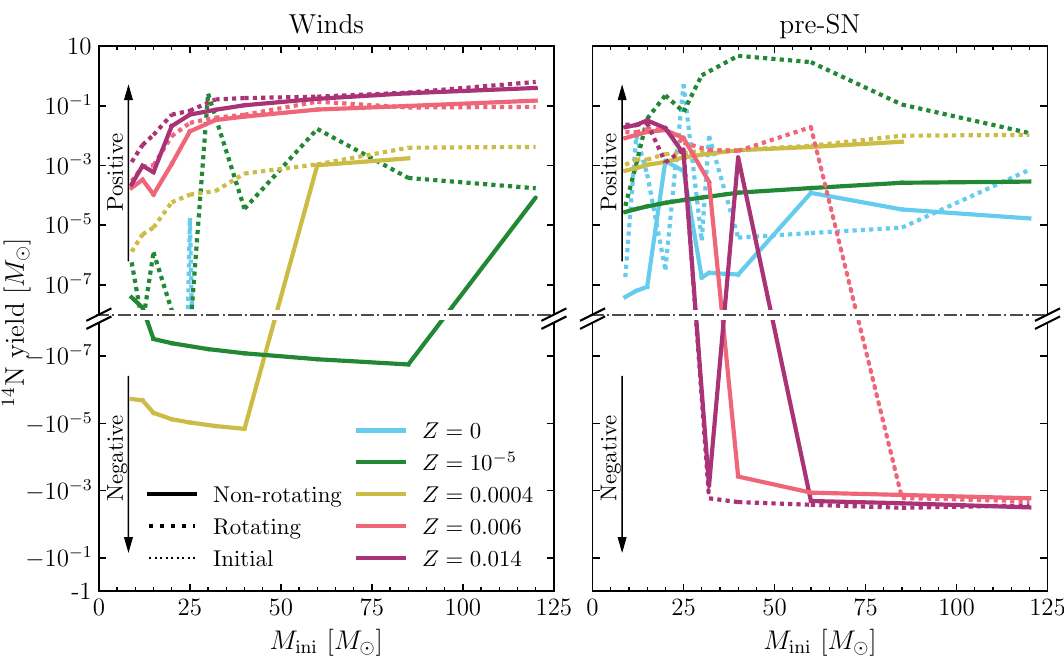}
 \caption{Stellar yields (in solar masses) of $^{14}$N ejected by winds during evolution (left panel, computed as in Eq.~\ref{eqn:N14_winds}) and integrated above the  mass coordinate of the CO core at the last computed stage of stellar evolution (right panel, computed as in Eq.~\ref{eqn:N14_preSN}) for all models with initial mass $M_{\rm ini}\geq9M_\odot$ at different metallicities ($Z=0$ (Pop~III, blue), $Z=10^{-5}$ (this paper, green), $Z=0.0004$ (I Zw 18, yellow), $Z=0.006$ (LMC, pink), $Z=0.014$ (solar, purple)). Because some models yield less $^{14}$N than they contain on the ZAMS, the yield can be negative. This is indicated on the figure by the separation in the vertical axes, the dash-dotted line, and the arrows showing `Positive' and `Negative'. The vertical axes are in different logarithmic scales. Solid lines: Non-rotating models. Dotted lines: Rotating models.}
 \label{N14_winds&SN_diffZ}
 \end{figure*}
Figure~\ref{Mfin_Mini_diffZ} shows the final mass $M_{\rm fin}$ (mass of the last computed model) as a function of the initial mass $M_{\rm ini}$ for all models with initial mass $M_{\rm ini}>\geq9M_\odot$ at different metallicities ($Z=0$ (Pop~III, blue), $Z=10^{-5}$ (this paper, green), $Z=0.0004$ (I Zw 18, yellow), $Z=0.006$ (LMC, pink), $Z=0.014$ (solar, purple)). Solid lines show non-rotating models and dotted lines, rotating ones.

Increasing the metallicity decreases the final mass of the models. This is due to the stronger stellar winds which incur greater mass-loss rates. At the low-end of the considered mass range, all stars follow the $M_{\rm fin}\simeq M_{\rm ini}$ line. The mass above which $M_{\rm fin} < M_{\rm ini}$ decreases as a function of metallicity: mass-loss becomes noticeable at lower initial masses for higher metallicities. Interestingly, the high-mass rotating $Z=10^{-5}$ models lose more mass than their $Z=0.0004$ counterparts. This is because they have larger luminosities during their entire evolution, and spend more time in the red part of the HRD.

For non-zero metallicity models (which in any case do not lose significant amounts of mass), rotation does not have a monotonous effect on the final mass: for low ($M_{\rm ini}<20-30\,M_\odot$ depending on $Z$) and high ($M_{\rm ini}>85\,M_\odot$) masses, rotation decreases the final mass, whereas for intermediate masses ($30\,M_\odot<M_{\rm ini}<85\,M_\odot$) it induces a larger final mass. This result is quite visible for $Z=10^{-5}$, $Z=0.006$, and $Z=0.014$, although the exact mass ranges vary with metallicity.\\

Figure~\ref{N14_winds&SN_diffZ} shows the stellar yields (in solar masses) of $^{14}$N for all models with initial mass $M_{\rm ini}\geq9M_\odot$ at different metallicities ($Z=0$ (Pop~III, blue), $Z=10^{-5}$ (this paper, green), $Z=0.0004$ (I Zw 18, yellow), $Z=0.006$ (LMC, pink), $Z=0.014$ (solar, purple)). We note that we define the stellar yield of an element as the ejected mass of this element minus its initial abundance in the star. The left panel of Fig.~\ref{N14_winds&SN_diffZ} shows the stellar yields of $^{14}$N ejected by winds, and the right panel shows the pre-supernova (pre-SN) yields. Because some models produce less $^{14}$N than they contain on the ZAMS, the yield can be negative. This is indicated on the figure by the separation in the vertical axes, the dash-dotted line, and the arrows showing ``Positive" and ``Negative". We note however that the total yields, taking into account both winds and pre-SN ejecta, are positive for all stars considered. In other words, when one is negative, it is always counterbalanced by the other. The positive and negative vertical axes are in logarithmic scale. Solid lines show non-rotating models and dotted lines, rotating ones. The thin dotted lines show the initial abundance of $^{14}$N in the models (when a thick curve overlaps with the thin one of the same colour, the mass of $^{14}$N ejected by that star is 0). 

The yields are computed in the following way. For winds, we multiply the mass lost (counted positively when mass is lost) by stars at each timestep by the surface mass fraction of nitrogen $x(^{14}{\rm N})_{\rm s}$, and sum this quantity over all timesteps of stellar evolution. This gives us the ejected mass, to which we subtract the product of the initial nitrogen abundance $x(^{14}{\rm N})_{\rm s, ini}$ by the total mass lost through winds $M_{\rm winds}$ in order to obtain the yield of nitrogen by winds $y(^{14}{\rm N})_{\rm winds}$:
\begin{equation}
\label{eqn:N14_winds}
    y(^{14}{\rm N})_{\rm winds} = \sum_{i=0}^{N}{(\delta M_{i}x(^{14}{\rm N})_{\rm s})} - (x(^{14}{\rm N})_{\rm s, ini}M_{\rm winds}),
\end{equation}
where $\delta M_{i}$ is the mass lost by the star at timestep $i$ and $N$ the total number of timesteps for that star.

For the pre-SN yields, we consider that all the nitrogen present above the CO core can be ejected. This is actually equal to the total quantity of nitrogen in the star at the pre-SN stage because there is no $\nuc{N}{14}$ inside the CO core. Since it is possible that stars do not eject their entire envelope, the yields that we provide are to be interpreted as an upper limit. At the last computed stage of stellar evolution, we integrate the $^{14}$N abundance in all the shells above the mass coordinate of the CO core, to obtain the ejected mass of $^{14}$N. In the case where the remnant is a white dwarf (WD) (mostly the 9\,$M_\odot$ models), the final event will be a planetary nebula (PN) and we give the $^{14}$N yield in the PN (we effectively treat it like a supernova). In this case, because nitrogen may be created dring thermal pulses of the AGB phase, the nitrogen yields we provide are lower bounds. We then subtract the product of the initial nitrogen abundance $x(^{14}{\rm N})_{\rm ini}$ by the mass ejected by the SN $M_{\rm SN}$ in order to obtain the pre-SN yield of nitrogen $y(^{14}{\rm N})_{\rm pre-SN}$:
\begin{equation}
\label{eqn:N14_preSN}
    y(^{14}{\rm N})_{\rm pre-SN} = \int_{M_{\rm CO}}^{M_{{\rm tot, f}}}{x(^{14}{\rm N})(m)\, \mathrm{d}m} - x(^{14}{\rm N})_{\rm ini}M_{\rm SN},
\end{equation}
where $x(^{14}{\rm N})(m)$ is the mass fraction of $^{14}$N in the shell at the mass coordinate $m$, $M_{\rm CO}$ is the mass of the CO core at the last computed stage, and $M_{{\rm tot, f}}$ is the final mass of the star. We of course have the relations $M_{\rm tot, f} = M_{\rm ini} - M_{\rm winds}$ and $M_{\rm tot, f} = M_{\rm CO} + M_{\rm SN}$.\\

The yield of nitrogen due to winds is more important at higher metallicities. This is expected because, as we mentioned previously, metallicity increases opacity and thus the amount of mass lost by winds. A larger metallicity also leads to a smaller minimal mass above which winds become important.\\
Rotation tends to increase mass loss as well as mixing efficiency (bringing more nitrogen to the surface), so it also increases the net $^{14}$N yield, and decreases the minimal mass for winds to be relevant (compared to the non-rotating models of the same metallicity). All these effects could be deduced from Fig.~\ref{Mfin_Mini_diffZ}.

One effect that appears in Fig.~\ref{N14_winds&SN_diffZ} is the larger yield of nitrogen for rotating $Z=10^{-5}$ models between 25 and 85\,$M_\odot$ than their $Z=0.0004$ counterparts (whereas they lose less mass). Indeed, the much more efficient mixing in the former (see Fig.~\ref{NHs_t_trel_diffZ_15M} and the discussion in Sect.~\ref{diffZ_surface}) leads to more nitrogen being brought up to the surface and ejected by the stellar winds. The 40\,$M_\odot$ model is the most striking example, and its high primary nitrogen production is also linked to the evolution of the hydrogen-burning shell (see below). Above 85\,$M_\odot$ however, the $Z=0.0004$ models eject more $^{14}$N than those at $Z=10^{-5}$. This is because, while the $Z=10^{-5}$ models lose more mass and see a larger relative enrichment in surface nitrogen, their absolute nitrogen surface abundances are much smaller than those of the $Z=0.0004$ models. As a result, even when subtracting the smaller initial $^{14}$N abundance of the EMP 85 and 120\,$M_\odot$ models, their yield through winds is smaller.

Most of the nitrogen ejected by the stars at higher (e.g. LMC and solar) metallicities is secondary nitrogen: it has been created through the CNO cycle from carbon and oxygen that were originally present in the star. For these stars, nitrogen production is thus proportional to their initial abundances of carbon and oxygen. In contrast, while nitrogen is also created by the same cycles in low-$Z$ stars, most of their carbon and oxygen has been produced by the triple-$\alpha$ and the $^{12}$C$(\alpha,\gamma)^{16}$O reaction during core helium-burning. As a result, nitrogen production does no longer scale with the initial metallicity. This primary nitrogen production depends on mixing processes between H- and He-burning regions, which are linked to convection and rotation. This last point would lead one to think that primary nitrogen production should be maximal for Pop~III stars. However, their convective cores are too small, such that the carbon produced in the helium-burning core does not reach the hydrogen-burning shell. Even if it could, the stars' surface temperatures are too high for them to develop large convective zones in order to efficiently transport the produced nitrogen to the surface. We would expect a more rapid rotation rate to better transport these chemical species and increase the nitrogen yield of these stars. Indeed, \citet{Tsiatsiou2024} computed fast-rotating ($\upsilon_{\rm ini}/\upsilon_{\rm crit}=0.7$) Pop~III models and found that these produce large amounts of primary nitrogen, comparable to the EMP models of the present study.\\

Table~\ref{tab:fates_diffZ} shows, for all massive ($M_{\rm ini}>9\,M_\odot$) stars at the five metallicities in this comparison, the final mass reached by each model, the mass of the CO core at the end of core helium-burning $M_{\rm CO}$, the mass fraction of carbon in the core $x(^{12}{\rm C})_{\rm c}$ at that same moment, and the type of remnant we expect. Some rows are filled with dashes because these specific models were not computed (e.g. some 30 and 32\,$M_\odot$ models). A thorough analysis of the fates of all the single star models computed with the \textsc{Genec} code is beyond the scope of this paper and will be the object of a dedicated forthcoming paper. Here we present just a few striking results.

In order to obtain the type of stellar remnant left behind, we follow the guidelines of \citet{Farmer2019} and \citet{Patton2020}. We look at the mass of the CO core $M_{\rm CO}$ (defined as the mass coordinate inside the star where the mass fraction of helium $x(^4{\rm He})<10^{-2}$) and the central mass fraction of carbon $x(^{12}{\rm C})_{\rm c}$ at the end of helium-burning. If $M_{\rm CO}<1.4\,M_\odot$, the remnant is a WD. If $1.4\,M_\odot<M_{\rm CO} <2.5\,M_\odot$, the remnant is a NS. If $2.5\,M_\odot<M_{\rm CO} <10\,M_\odot$, we look at Fig.~3 of \citet{Patton2020} to determine if the star will implode or explode (this depends on $x(^{12}{\rm C})_{\rm c}$ at the end of helium-burning). If the star explodes, the remnant is a NS; if it implodes, or if $10\,M_\odot<M_{\rm CO}<38\,M_\odot$, the remnant is a BH. If $M_{\rm CO}>38\,M_\odot$, we find the star with the closest CO core mass in the table provided by \citet{Farmer2019} to obtain the outcome (pulsational pair-instability supernova with a BH remnant or pair-instability supernova with no remnant).

For non-rotating models with initial mass $M_{\rm ini}\geq20\,M_\odot$, the mass of the CO core tends to be a decreasing function of metallicity. This is due to the metallicity dependence on mass loss: indeed, for these models the final mass is also a decreasing function of metallicity, and this has an impact on $M_{\rm CO}$.\\
A notable exception is the non-rotating 20\,$M_\odot$ model at EMP metallicity that has the second-lowest $M_{\rm CO}$ of the non-rotating 20\,$M_\odot$ models. We predict its remnant to be a BH, although this is to be taken with a grain of salt as the model resides in a part of the $(M_{\rm CO}, x(^{12}{\rm C})_{\rm c})$ parameter space where very small variations in either $M_{\rm CO}$ or $x(^{12}{\rm C})_{\rm c}$ can change the fate of the star \citep[see Fig.~3 of][]{Patton2020}. This is also the case with the rotating 20\,$M_\odot$ Pop~III and the non-rotating 25\,$M_\odot$ $Z=0.0004$ models.\\
For the rotating 25\,$M_\odot$ model at solar metallicity and the rotating 32\,$M_\odot$ model at $Z=0.006$, even though their $M_{\rm CO}$ and $x(^{12}{\rm C})_{\rm c}$ are similar to those of their counterparts at other metallicities, they are predicted to leave a NS and not a BH. Looking at Fig.~3 of \citet{Patton2020}, we can see that they fall in pockets of explodability, and we consider this result to be reliable.\\
Another model presenting an interesting difference with its counterparts at other metallicities is the rotating EMP 40\,$M_\odot$ model, whose CO core mass is $M_{\rm CO}=3.14\,M_\odot$, much lower than the other models, and for which we predict its remnant to be a NS. Compared to the other models at the end of core helium-burning, this star has a very extended intermediate convective zone associated with the hydrogen-burning shell, and this shell extends down to a mass coordinate of 5.5\,$M_\odot$ inside the star (against 14.4\,$M_\odot$ for the Pop~III model and 17\,$M_\odot$ for the $Z=0.0004$ one). This position of the hydrogen-burning shell prevents the core from being as large as in other models, explaining the small $M_{\rm CO}$ for this model. Also (this is the case for the rotating 30 and 60\,$M_\odot$ models as well), the inwards movement of the hydrogen-burning shell during core helium-burning makes it pass through regions that have previously been enriched in carbon and oxygen by the fusion of helium in the core (which has now receded). This leads the CNO cycle to produce a large quantity of $^{14}$N in the shell and is the reason behind the very large $^{14}$N yields for these stars.\\

\section{Comparison with previous works} \label{comparisons}

Most theoretical works on EMP stars have focused on low to intermediate-mass stars \citep[see e.g.][]{Herwig2004,Hirschi2007,GilPons2013,GilPons2021,Ventura2021}, which may still be alive today in our galaxy. There has however been a few studies investigating massive stars at different extremely low metallicities (including $Z\sim10^{-5}$), such as \citet{Limongi2018}. In this section we concentrate on the results of \citet{GilPons2013} and \citet{GilPons2021} for low to intermediate-mass stars and \citet{Limongi2018} for massive stars. We also compare our results to a grid of models at $Z=10^{-5}$ computed by Costa \& Shepherd (in prep.).\\

\begin{figure*}[h!]
    \centering
    \includegraphics{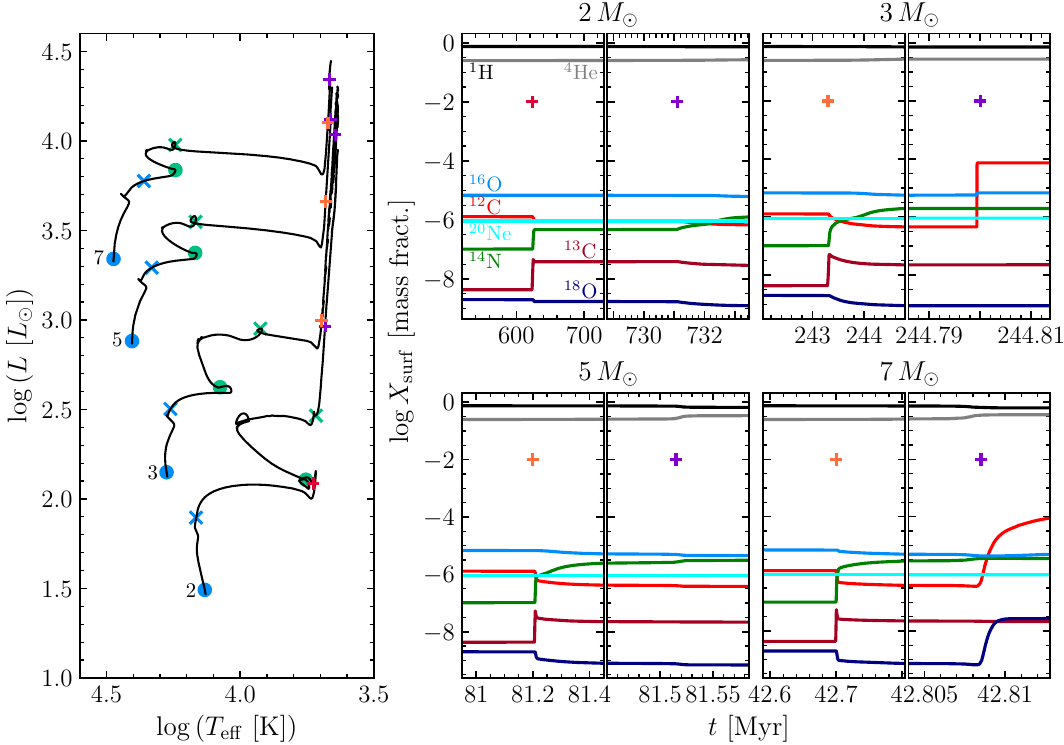}
    \caption{Dredge-up properties of the non-rotating 2, 3, 5, and 7\,$M_\odot$ models. Left panel: HRD of the four models. The blue and green circles and crosses show the beginning and end of core hydrogen and core helium burning. The red, orange, and purple plus signs show the moments of the first (2\,$M_\odot$), second (3, 5, and 7\,$M_\odot$), and third dredge-up episodes respectively. Right panels: surface abundances around the two dredge-up episodes that each star experiences. The initial mass of the corresponding model is indicated at the top of each subpanel.}
    \label{fig:HRD_DU}
\end{figure*}
\begin{figure*}[h!]
    \centering
    \includegraphics{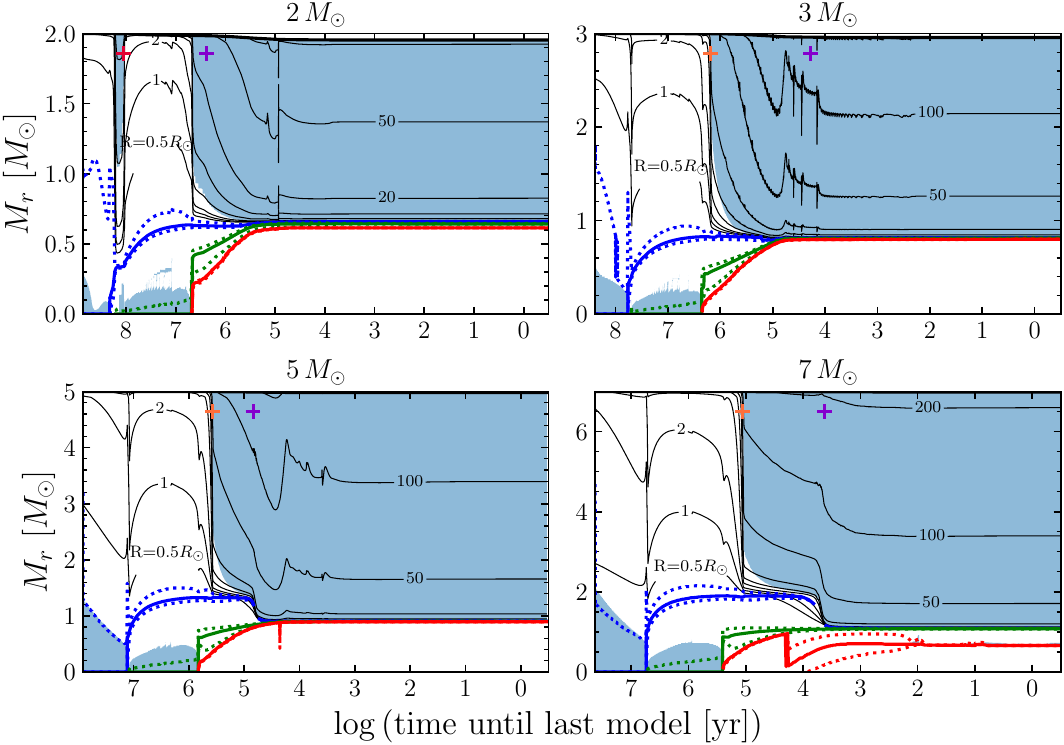}
    \caption{Kippenhahn diagrams of the non-rotating 2, 3, 5, and 7\,$M_\odot$ models. The blue, green, and red solid lines show the location of maximum nuclear energy generation for hydrogen, helium, and carbon burning respectively. Energy is generated in the region delimited by the dotted lines. The red, orange, and purple plus signs show the moments of the first (2\,$M_\odot$), second (3, 5, and 7\,$M_\odot$), and third dredge-up episodes respectively. The black lines are isoradius lines. The initial mass of the corresponding model is indicated at the top of each panel.}
    \label{fig:Kipp_DU}
\end{figure*}

\citet{GilPons2013} and \citet{GilPons2021} used the Monash University Stellar Evolution code MONSTAR to compute the evolution of non-rotating intermediate-mass stars at $Z=10^{-5}$ up to the late thermally pulsing (super) asymptotic giant branch (TP-(S)AGB) phase. \citet{GilPons2013} focused on the stellar evolution and \citet{GilPons2021} on the nucleosynthetic yields. Compared to \citet{GilPons2013}, our models have longer MS and shorter core He-burning phases, though they have similar total lifetimes. For instance, the 4\,$M_\odot$ model of \citet{GilPons2013} has $\tau_{\rm H}=96.2$\,Myr and $\tau_{\rm He}=28.2$\,Myr (total lifetime of 124.4\,Myr), while ours has $\tau_{\rm H}=103.9$\,Myr and $\tau_{\rm He}=20.4$\,Myr (total lifetime of 124.3\,Myr). Nevertheless, the tracks in the HRD are remarkably similar.

Figure~\ref{fig:HRD_DU} shows the HRD (left panel) and surface abundance evolution (right panels) during the dredge-up (DU) episodes for the non-rotating 2, 3, 5, and 7\,$M_\odot$ models we computed. We do not show the 1.7, 2.5, and 4\,$M_\odot$ stars because they behave in a similar manner to the 2, 3, and 5\,$M_\odot$ models respectively. Figure~\ref{fig:Kipp_DU} shows the Kippenhahn diagrams of the same stars. In both figures, the red, orange, and purple plus signs show the moments of the first (2\,$M_\odot$ only), second (SDU, 3, 5, and 7\,$M_\odot$), and third (TDU) dredge-up episodes respectively. In comparing our Fig.~\ref{fig:HRD_DU} to Figs.~3-6 of \citet{GilPons2021}, it is essential to note that we do not compute the TP-(S)AGB phase. This is why the subpanels annotated with the purple plus signs in Fig.~\ref{fig:HRD_DU} show very different behaviours than the corresponding subpanels in Figs.~3-6 of \citet{GilPons2021}.

For the second dredge-up episode (panels annotated with the orange plus signs in Fig.~\ref{fig:HRD_DU}), our surface abundances evolve in a similar manner to those of \citet{GilPons2021}: the abundance of $^{12}$C decreases while those of $^{13}$C and $^{14}$N increase, and that of $^{16}$O does not vary much. For both the 3 and 7\,$M_\odot$ models, we find that the surface abundance of $^{14}$N is larger than that of $^{12}$C after the SDU. The surface abundance of $^{16}$O remains larger than that of $^{14}$N in our models \citep[contrary to the models of][especially their 7\,$M_\odot$ one]{GilPons2021}, which we attribute to the $\alpha$-enhanced initial composition of our models. The main difference in surface abundance evoution is for $^{18}$O, which decreases in our case and increases for the models of \citet{GilPons2021}.

The TDU episodes present much more marked differences. In our case, the surface abundance of $^{14}$N increases only marginally whereas its increase is the strongest of all elements followed by \citet{GilPons2021}. Overall, our models produce much more carbon-enhanced and much less nitrogen-enhanced stars than those of \citet{GilPons2021}.\\

\citet{Limongi2018} computed models of massive stars between 13 and 120\,$M_\odot$ at four metallicities $({\rm [Fe/H]}=-3, -2, -1, 0)$ and three initial rotation velocities $(\upsilon_{\rm ini}=0, 150\,{\rm km\,s}^{-1}, 300\,{\rm km\,s}^{-1})$ using the updated \textsc{FRANEC} code \citep[first introduced in][]{Chieffi2013}. Here we focus our comparison on models at ${\rm [Fe/H]}=-3$, corresponding to $Z=3.236\times10^{-5}$. Since they use the same initial rotation velocity for all masses, there is no direct equivalence between their models and ours at $\upsilon_{\rm ini}/\upsilon_{\rm crit}=0.4$; however, the models at $\upsilon_{\rm ini}=300\,{\rm km\,s}^{-1}$ correspond to values of $\upsilon_{\rm ini}/\upsilon_{\rm crit}$ between 0.2 and 0.4 for masses between 12 and 120\,$M_\odot$. Thus, we consider this set for our comparison of rotating models, keeping in mind that the impact of rotation-induced effects should be smaller for the more massive models of \citet{Limongi2018}.

The overall evolution of their non-rotating stars is similar to our models, up to the end of core helium-burning. The main difference until then occurs during the beginning of core helium-burning, where the models of \citet{Limongi2018} evolve bluewards before becoming cooler (this is most noticeable in their models between 13 and 25\,$M_\odot$). Our models exhibit a similar behaviour, but this bluewards evolution is less marked. In the grid of \citet{Limongi2018}, stars between 13 and 25\,$M_\odot$ end their evolution as RSG. In contrast, only the 12, 60, and 85\,$M_\odot$ non-rotating stars end their lives as RSG in our models. In both grids, stars exhaust helium in the core as BSG and evolve redwards only later.

The rotating models of \citet{Limongi2018} reach the end of core helium-burning as RSG ($M_{\rm ini}\leq25\,M_\odot$, except for the 20\,$M_\odot$ star) or WR ($M_{\rm ini}\geq30\,M_\odot$). Some of them reach the WR phase even before core helium ignition. In contrast, most of our rotating stars (except for the 20, 25, and 40\,$M_\odot$ ones) finish burning helium in the core as RSG. Even though the rotating 30, 60, 85, and 120\,$M_\odot$ models in our grid end their evolution with a rather low surface hydrogen mass fraction ($X_{\rm s}<0.3$), they are too cold to be classified as WR stars (for which a typical criterion is $\log{(T_{\rm eff}~[K])}>4$).

Looking at cores, \citet{Limongi2018} find that their rotating stars have more massive CO cores than the non-rotating ones do, by about 50\,\%. In our case, $M_{\rm CO}$ is smaller for rotating models below and including 60\,$M_\odot$, and larger for the rotating 85 and 120\,$M_\odot$ models than for the non-rotating ones. The abundance of carbon in the core at the end of helium burning (which, coupled with the mass of the CO core, is useful in order to determine the final fate of stars with $M_{\rm CO}<10\,M_\odot$, see Sect.~\ref{diffZ_final}) does not vary much between our rotating and non-rotating models. \citet{Limongi2018} find a large discrepancy, with rotating models being much poorer in carbon than non-rotating ones (see their Fig.~19). They explain this discrepancy by the effect of rotational mixing which brings fresh helium from the hydrogen-burning shell into the helium-burning core, driving the $^{12}{\rm C}(\alpha,\gamma)^{16}{\rm O}$ reaction. The fact that the CO cores of our rotating models are not more massive than those of the non-rotating ones may lead to a larger separation between the boundary of the helium-burning core and the bottom of the hydrogen-burning shell, making the aforementioned rotation-induced mixing less effective.

Overall, it is difficult to compare our rotating models with those of \citet{Limongi2018}, because the different choice of initial rotation makes it apples to oranges. Nevertheless, the differences we highlighted above are significant enough to be noted. A part of these differences is due to the different way the effects of rotation are implemented into the two grids models \citep[see the discussion in][]{Nandal2023}.\\

\begin{figure*}[h!]
    \centering
    \includegraphics{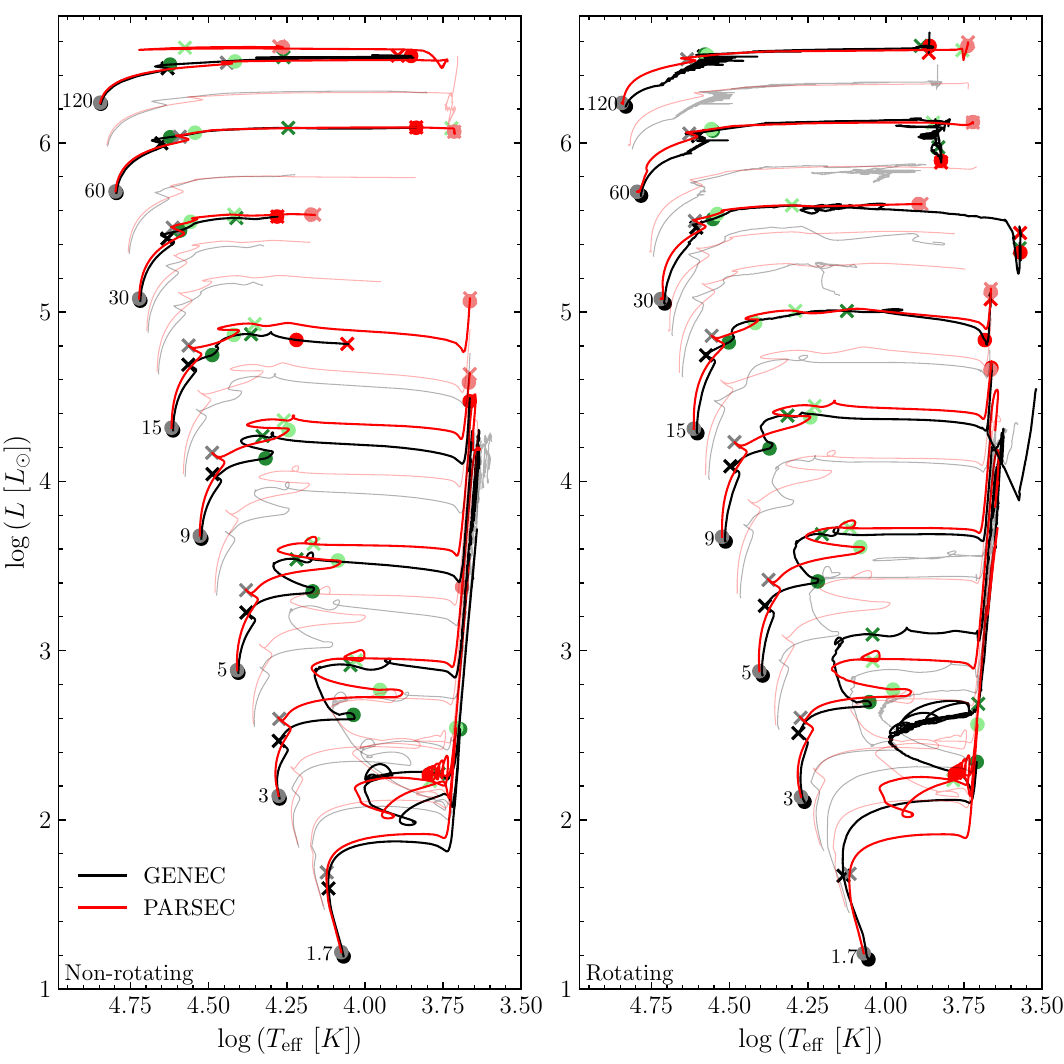}
    \caption{HRD comparing the models presented in the present paper (black lines) and a grid of models computed with \textsc{parsec} by Costa \& Shepherd (in prep., red lines). We select a few representative initial masses (1.7, 3, 5, 9, 15, 30, 60, 120\,$M_\odot$) for which we show the beginning (circles) and end (crosses) of the core hydrogen (black), helium (green), and carbon (red) burning phases. The other masses are shown with a lower opacity for increased clarity.}
    \label{fig:HRD_PARSEC}
\end{figure*}

The PAdova and tRieste Stellar Evolutionary Code \citep[\textsc{parsec},][]{Bressan2012}{}{} has been utilised to produce large grids of stellar evolutionary tracks and isochrones.
\textsc{parsec} follows the evolution of single stars from the pre-MS to the most advanced burning phases. Other main input physics used in \textsc{parsec} V2.0 are described thoroughly in \citet{Bressan2012}, \citet{Chen2014}, \citet{Fu2018}, \citet{Costa2019}, and \citet{Nguyen2022}. We compare our models with a new grid computed at $Z=10^{-5}$ for Costa \& Shepherd (in prep.).\\

Figure~\ref{fig:HRD_PARSEC} shows the HRD of the two sets of models computed with the \textsc{Genec} (black) and \textsc{parsec} (red) codes. We mark the beginning (circles) and end (crosses) of the hydrogen (black), helium (green), and carbon (red) burning phases on the tracks for a few representative masses (1.7, 3, 5, 9, 15, 30, 60, 120\,$M_\odot$), and the other masses are shown with a lower opacity for increased clarity.

Models start the MS at nearly identical locations on the HRD for both rotating and non-rotating cases. The MS tracks for non-rotating stars are nearly identical below 2\,$M_\odot$ and above 30\,$M_\odot$. Between 2.5 and 25\,$M_\odot$ however, the \textsc{parsec} models are consistently brighter than their \textsc{Genec} counterparts.

This difference in luminosity carries on at core helium ignition. Throughout the helium-burning phase, both models evolve redwards and end that phase at a similar $T_{\rm eff}$, although the luminosity discrepancy remains. This can be seen for instance in the blue loops, which occur for the same masses with both codes, but at higher luminosity for stars computed with \textsc{parsec}.

Following core helium burning, the main difference can be seen in models above and including 15\,$M_\odot$. Apart for the 120\,$M_\odot$ one, \textsc{parsec} models consistently experience a stronger expansion than our models, and ignite carbon in the core at a much lower effective temperature. This is most prominently the case for the 15\,$M_\odot$ stars, where the one computed with \textsc{parsec} begins central carbon burning as a RSG, having developed an extended convective envelope, whereas the \textsc{Genec} model remains hotter and more compact, and completes central carbon burning in degenerate conditions as a BSG. A noteworthy difference between the two is their CO core masses at the end of core helium burning: the \textsc{parsec} model has $M_{\rm CO}=3.16\,M_\odot$ and ours has $M_{\rm CO}=2.26\,M_\odot$.

This discrepancy in CO core mass is present for almost all models. We note as well that the \textsc{parsec} models are similar in that respect to those of \citet{Limongi2018} discussed above: in both cases, stars boast much more massive CO cores (\textsc{parsec} CO cores are approximately 40\,\% larger for non-rotating models and 50\,\% larger for rotating models compared to our models) at the end of helium burning. Both evolve to much cooler effective temperatures before igniting carbon, than our models.

During core helium burning, both 120\,$M_{\odot}$ models evolve to $\log{(T_{\rm eff})}\sim 3.75$ and subsequently evolve bluewards before completing central helium burning. Our model crosses the HRD once more before igniting and depleting carbon in the core, while the \textsc{parsec} model undergoes carbon burning shortly before becoming unstable to pair instability, when the model is stopped.

Looking at surface nitrogen enrichment, we find that for stellar masses below 15\,$M_\odot$, both \textsc{parsec} and \textsc{Genec} models exhibit concordant final surface nitrogen abundances. However, for masses greater than 15\,$M_\odot$, non-rotating \textsc{parsec} models exhibit larger enhancement in surface nitrogen than ours. The difference is likely due to \textsc{Genec} models never getting cool enough to develop a convective envelope. As a result they keep their initial surface composition throughout their evolution.
For rotating models, \textsc{Genec} models show much more surface nitrogen than the \textsc{parsec} models; this is a consequence of the different treatments of rotation-induced mixing. 
Non-rotating stars below 60 M$_{\odot}$ see almost no change in their surface hydrogen for both \textsc{parsec} and \textsc{Genec} models. Above 60\,$M_{\odot}$, \textsc{parsec} models show decreased surface hydrogen, which occurs only in our 120\,$M_\odot$ model. In the case of rotating stars, all models see a decline in their surface hydrogen abundance, although this decline is stronger for \textsc{parsec} models than for the \textsc{Genec} ones. Part of the differences mentioned are linked to the different treatments of core overshoot and of the effects of rotation.

\section{Discussion and conclusions} \label{Discussion_conclusion}

In this paper, we introduce the last of the series of \textsc{Genec} grids of stellar models with rotation, which use the same physical ingredients as those described in \citet{Z0142012}. The aim of this approach is to provide homogeneous grids covering a wide range of metallicities. The obvious drawback to using these grids is that we do not incorporate the latest results on mass-loss rates, the physics of rotation, or the transport of angular momentum or convection (to name just a few).

As is the case for all previous papers in this series, we consider only the case of single stars. As such, our results do not apply to stars in binaries (unless the orbital separation is too large for the binary interactions to play a role in stellar evolution).
Also, we do not take into account magnetic fields. New prescriptions for the transport of angular momentum by internal magnetic fields \citep{Eggenberger2022} based on the revised version of the Tayler-Spruit dynamo \citep{Spruit2002,Fuller2019} have been implemented in the \textsc{Genec} code \citep[see e.g. ][]{Moyano2023,Nandal2023} and these new physics will be included in the next series of grids. While being able to better reproduce the internal rotation of the Sun \citep{egg2022b}, of solar-type stars \citep{bet2023}, of gamma Doradus pulsators \citep{Moyano2023}, and of red giants \citep{Fuller2019}, and the natal spins of neutron stars \citep[see e.g.][]{Ma2019}, we remind the reader that these magnetic models are however not able to provide a global solution to the internal transport of angular momentum using a unique value for the calibration parameter entering this formalism for stars at different evolutionary stages \citep[][]{egg2019,den2020,moy2024}.

Our results regarding the fates of massive stars are contingent on the choice of recipe used to predict the remnant type. Although we use and trust the results from \citet{Farmer2019} and \citet{Patton2020}, they have their limitations: \citet{Farmer2019} compute the evolution of helium stars and \citet{Patton2020} of naked CO cores. While the assumption of the most massive stars becoming helium stars makes sense due to the strong mass losses involved in these stars, it may not be the most justified in the extremely metal-poor models presented here. Indeed, they still contain hydrogen at the pre-SN stage, which may influence their fate. Whether a massive star will implode or explode at the end of its evolution has a big impact on the surrounding medium and, to a larger extent, on galactic chemical evolution. Nevertheless, this question remains uncertain. Recent works predicting the fates of massive stars include \citet{Ertl2016}, \citet{Muller2016}, \citet{Boccioli2023}, and \citet{Schneider2024}, among others.\\

We summarise our main results below:
\begin{itemize}
    \item We do not find any WR stars at this extremely low metallicity.
    \item In terms of evolution in the HRD, the present models at $Z=10^{-5}$ fit well between those at surrounding metallicities ($Z=0$ and $Z=4\times10^{-4}$). When comparing models at the same evolutionary stage, the present models tend to be hotter and more luminous than the stars at $Z=4\times10^{-4}$, and colder and less luminous than those at $Z=0$.
    \item Stellar lifetime is a decreasing function of metallicity for initial masses below 10\,$M_\odot$. For larger initial masses, lifetime depends much less on metallicity.
    \item Stars at lower metallicities tend to lose less mass and thus end their evolution with larger final masses than their counterparts at higher metallicities. Mass-loss processes only become relevant at the present metallicity for the most massive stars ($M_{\rm ini}\geq85-120\,M_\odot$)
    \item  Of all the metallicities studied in this series of papers, rotating stars at $Z=10^{-5}$ are the most efficient producers of primary nitrogen. While stars at higher metallicities eject more (secondary) nitrogen during their evolution through stellar winds, the present models are the largest producers of nitrogen at the pre-SN stage. 
    \item In general, our models have a lower luminosity and less massive CO cores than the same stars computed with other stellar evolution codes (such as \textsc{parsec}), and our non-rotating massive stars between 15 and 40\,$M_\odot$ spend their post-MS evolution at larger effective temperatures. This is likely due to different choices of physics (such as overshooting, mass-loss, and opacity, among others).
\end{itemize}

\begin{table*}
    \centering
    \caption{Final properties and fates of massive stars at five different metallicities.}
    \begin{tabular}{cc|cccc|cccc}
    \hline\hline
    $M_{\rm ini}$ & $Z$ & \multicolumn{4}{c|}{Non-rotating} & \multicolumn{4}{c}{Rotating} \\
    & & $M_{\rm fin}$ & $M_{\rm CO}$ & $x(^{12}{\rm C})_{\rm c}$ & Fate & $M_{\rm fin}$ & $M_{\rm CO}$ & $x(^{12}{\rm C})_{\rm c}$ & Fate \\
    $M_\odot$ & & \multicolumn{2}{c}{$M_\odot$} & mass fract. & & \multicolumn{2}{c}{$M_\odot$} & mass fract. & \\
    \hline
    \multirow{5}{*}{9}& 0 & 8.8 & 0.72 & 0.40 & WD &8.99 & 0.90 & 0.32 & WD \\
    & $10^{-5}$ & 8.80 & 1.01 & 0.40 & WD &8.76 & 1.12 & 0.35 & WD \\
    & 0.0004 & 8.79 & 1.05 & 0.40 & WD &8.78 & 1.15 & 0.28 & WD \\
    & 0.006 & 8.44 & 0.88 & 0.38 & WD &8.57 & 1.00 & 0.37 & WD \\
    & 0.014 & 8.59 & 0.83 & 0.40 & WD &8.35 & 1.18 & 0.29 & WD \\
    \hline\multirow{5}{*}{12}& 0 & 11.76 & 1.45 & 0.36 & NS &11.76 & 1.74 & 0.15 & NS \\
    & $10^{-5}$ & 11.74 & 1.57 & 0.36 & NS &11.76 & 1.59 & 0.30 & NS \\
    & 0.0004 & 11.69 & 1.70 & 0.36 & NS &11.67 & 1.80 & 0.33 & NS \\
    & 0.006 & 10.66 & 1.46 & 0.35 & NS &10.95 & 1.77 & 0.27 & NS \\
    & 0.014 & 11.08 & 1.26 & 0.35 & WD &10.02 & 1.82 & 0.25 & NS \\
    \hline\multirow{5}{*}{15}& 0 & 14.70 & 2.20 & 0.34 & NS &15.00 & 1.81 & 0.33 & NS \\
    & $10^{-5}$ & 14.69 & 2.26 & 0.34 & NS &14.80 & 2.64 & 0.23 & NS \\
    & 0.0004 & 14.62 & 2.38 & 0.33 & NS &14.90 & 2.56 & 0.22 & NS \\
    & 0.006 & 13.62 & 2.30 & 0.34 & NS &13.75 & 2.69 & 0.28 & NS \\
    & 0.014 & 12.98 & 2.11 & 0.36 & NS &10.85 & 2.68 & 0.20 & NS \\
    \hline\multirow{5}{*}{20}& 0 & 19.58 & 4.05 & 0.30 & NS &20.00 & 3.98 & 0.17 & NS \\
    & $10^{-5}$ & 19.60 & 3.75 & 0.29 & BH &19.98 & 2.98 & 0.26 & BH \\
    & 0.0004 & 19.56 & 3.97 & 0.25 & NS &18.98 & 4.84 & 0.23 & BH \\
    & 0.006 & 14.63 & 3.93 & 0.26 & NS &13.02 & 4.33 & 0.22 & BH \\
    & 0.014 & 8.46 & 3.65 & 0.31 & NS &7.18 & 4.36 & 0.27 & BH \\
    \hline\multirow{5}{*}{25}& 0 & 24.50 & 5.89 & 0.27 & NS &24.93 & 3.00 & 0.25 & BH \\
    & $10^{-5}$ & 24.50 & 5.52 & 0.26 & NS &24.97 & 5.51 & 0.23 & BH \\
    & 0.0004 & 24.41 & 5.67 & 0.27 & BH &23.62 & 6.62 & 0.21 & BH  \\
    & 0.006 & 11.22 & 5.68 & 0.26 & NS &11.47 & 6.27 & 0.22 & BH  \\
    & 0.014 & 8.12 & 5.29 & 0.28 & NS &9.69 & 6.53 & 0.25 & NS \\
    \hline\multirow{5}{*}{30}& 0 & 29.40 & 8.21 & 0.26 & BH  &30.00 & 6.42 & 0.17 & BH  \\
    & $10^{-5}$ & 29.39 & 7.54 & 0.25 & BH  &21.61 & 4.80 & 0.21 & BH \\
    & 0.0004 & -- & -- & -- & -- & -- & -- & -- & -- \\
    & 0.006 & -- & -- & -- & -- & -- & -- & -- & -- \\
    & 0.014 & -- & -- & -- & -- & -- & -- & -- & -- \\
    \hline\multirow{5}{*}{32}& 0 & 31.36 & 8.86 & 0.23 & BH &31.97 & 6.50 & 0.14 & BH \\
    & $10^{-5}$ & -- & -- & -- & -- & -- & -- & -- & -- \\
    & 0.0004 & 31.14 & 8.46 & 0.24 & BH &29.93 & 8.75 & 0.20 & BH \\
    & 0.006 & 11.57 & 8.36 & 0.22 & BH &13.61 & 9.91 & 0.22 & NS \\
    & 0.014 & 10.70 & 7.70 & 0.26 & BH &10.12 & 7.03 & 0.24 & BH \\
    \hline\multirow{5}{*}{40}& 0 & 39.20 & 12.50 & 0.20 & BH &40.00 & 10.24 & 0.31 & BH \\
    & $10^{-5}$ & 39.19 & 11.67 & 0.21 & BH &39.79 & 3.14 & 0.31 & NS \\
    & 0.0004 & 37.07 & 11.63 & 0.22 & BH &33.89 & 12.52 & 0.17 & BH \\
    & 0.006 & 14.97 & 11.51 & 0.19 & BH &18.92 & 14.89 & 0.19 & BH \\
    & 0.014 & 13.73 & 10.37 & 0.22 & BH &12.33 & 8.98 & 0.23 & BH \\
    \hline\multirow{5}{*}{60}& 0 & 58.80 & 22.57 & 0.16 & BH &59.70 & 19.37 & 0.15 & BH \\
    & $10^{-5}$ & 58.78 & 20.92 & 0.16 & BH &56.39 & 13.42 & 0.18 & BH \\
    & 0.0004 & 44.79 & 20.63 & 0.17 & BH &49.81 & 22.93 & 0.14 & BH \\
    & 0.006 & 22.45 & 18.15 & 0.14 & BH &32.78 & 28.19 & 0.13 & BH \\
    & 0.014 & 12.25 & 9.14 & 0.23 & BH &17.98 & 13.96 & 0.20 & BH \\
    \hline\multirow{5}{*}{85}& 0 & 83.09 & 31.77 & 0.12 & BH &83.98 & 30.46 & 0.18 & BH \\
    & $10^{-5}$ & 82.76 & 31.62 & 0.14 & BH &55.42 & 37.36 & 0.09 & BH \\
    & 0.0004 & 64.72 & 32.33 & 0.13 & BH &57.83 & 40.05 & 0.08 & BH \\
    & 0.006 & 30.83 & 26.04 & 0.15 & BH &35.81 & 29.91 & 0.14 & BH \\
    & 0.014 & 18.28 & 14.65 & 0.19 & BH &26.39 & 21.44 & 0.16 & BH \\
    \hline\multirow{5}{*}{120}& 0 & 118.32 & 52.89 & 0.10 & PPISN &116.44 & 56.40 & 0.27 & PPISN \\
    & $10^{-5}$ & 94.85 & 47.98 & 0.05 & PPISN &84.56 & 59.44 & 0.08 & PISN \\
    & 0.0004 & -- & -- & -- & -- &92.29 & 61.26 & 0.07 & PISN \\
    & 0.006 & 53.20 & 47.14 & 0.10 & PPISN &52.44 & 45.07 & 0.11 & PPISN \\
    & 0.014 & 30.29 & 25.48 & 0.15 & BH &19.04 & 14.86 & 0.19 & BH \\
    \hline
    \end{tabular}
    \label{tab:fates_diffZ}
\end{table*}

\begin{acknowledgements}
YS, ST, DN, FDM, CG, SE, PE, and GM have received funding from the European Research Council (ERC) under the European Union's Horizon 2020 research and innovation programme (grant agreement No 833925, project STAREX). KGS acknowledges funding support by the Italian Ministerial Grant PRIN 2022, “Radiative opacities for astrophysical applications”, No. 2022NEXMP8, CUP G53D23000910006. NY acknowledges the Fundamental Research Grant Scheme grant number FRGS/1/2018/STG02/UM/01/2 and FRGS/1/2021/STG07/UM/02/4 under Ministry of Higher Education, Malaysia. RH acknowledges support from STFC, the World Premier International Research Centre Initiative (WPI Initiative), MEXT, Japan, the IReNA AccelNet Network of Networks (National Science Foundation, Grant No. OISE-1927130) and the European Union’s Horizon 2020 research and innovation programme (ChETEC-INFRA, Grant No. 101008324). LS has received support from the SNF project No 212143. JB and GB acknowledge funding by the SNF AMBIZIONE grant No 185805 (Seismic inversions and modelling of transport processes in stars).
\end{acknowledgements}

\bibliographystyle{aa}
\bibliography{References}

\end{document}